\documentclass[proof]{rmaa}
\usepackage[latin1]{inputenc}
\usepackage{paralist}

\title{New insights into the nature of the SMC WR/LBV binary HD 5980\footnote{Based on observations made with the NTT, MPI 2.2m and Danish 1.5m telescopes at ESO, La Silla, Chile}
}
\author{C. Foellmi\altaffilmark{1},         
        G. Koenigsberger\altaffilmark{2},     
        L. Georgiev\altaffilmark{3},          
        O. Toledano\altaffilmark{2}           
        S.V. Marchenko \altaffilmark{4},       
        P. Massey\altaffilmark{5},        
        T. H. Dall \altaffilmark{6},          
        A.F.J. Moffat\altaffilmark{7},    
        N. Morrell\altaffilmark{8},       
        M. Corcoran\altaffilmark{9,10},        
        A. Kaufer\altaffilmark{11},       
        Y. Naz\'e\altaffilmark{12},       
        J. Pittard\altaffilmark{13},      
        N. St.-Louis\altaffilmark{7},     
        A. Fullerton\altaffilmark{14},     
        D. Massa\altaffilmark{15}         
        A. M. T. Pollock\altaffilmark{17} 
      }       
\altaffiltext{1}{Laboratoire d'AstrOphysique de Grenoble, 414 Rue de la Piscine, Domaine Universitaire, 38400 Saint-Martin d'H\`eres, France}              
\altaffiltext{2}{Instituto de Ciencias F\'{\i}sicas, Univer\-sidad Nacio\-nal Aut\'o\-noma de M\'exico, Cuernavaca, Mor., Mexico}
\altaffiltext{3}{Instituto de Astronom\'{\i}a, Univer\-sidad Nacio\-nal Aut\'o\-noma de M\'exico, M\'exico D.~F. 04510}
\altaffiltext{4}{Department of Physics and Astronomy, Western Kentucky University, 1906 College Heights Blvd \#11001, Bowling Green, KY 42101-1101, USA}
\altaffiltext{5}{Lowell Observatory, 1400 W Mars Hill Rd., Flagstaff, AZ, 86001,USA}   
\altaffiltext{6}{Gemini Observatory, 670 N. A'ohoku Pl., Hilo, HI 96720, USA}
\altaffiltext{7}{ D\'epartement de Physique, Universit\'e de Montr\'eal, C.P. 6128, Succ. C-V, Montr\'eal, QC, H3Y 1C5, and Observatoire du mont M\'egamtic, Canada}
\altaffiltext{8}{Las Campanas Observatory, Carnegie Observatories, Casilla 601, La Serena, Chile}
\altaffiltext{9}{CRESST and X-ray Astrophysics Laboratory NASA/GSFC, Greenbelt, MD 20771, USA.}
\altaffiltext{10}{Universities Space Research Association, 10211 Wincopin Circle, Suite 500 Columbia, MD 21044, USA.}
\altaffiltext{11}{European Southern Observatory, Alonso de Cordova 3107, Casilla 19001, Santiago 19, Chile}  
\altaffiltext{12}{FNRS and Institut d'Astrophysique et de G\'eophysique, Universit\'e de Li\`ege, All\'ee du 6 Ao\^ut 17, Bat B5C, B4000-Li\`ege, Belgium}
\altaffiltext{13}{School of Physics and Astronomy, University of Leeds, Leeds, UK}
\altaffiltext{14}{Space Telescope Science Institute, 3700 San Martin Drive, Baltimore, MD 21218}     
\altaffiltext{15}{NASA/GSFC, Mailstop 665.0, Greenbelt, MD 20771}       
\altaffiltext{17}{European Space Agency, XMM-Newton SOC, European Space Astronomy Centre, Apartado 78, 28691 Villanueva de la Caqada, Madrid, Spain.}        
 
\shortauthor{Foellmi et~al.}
\Note{Version: 12 September 2007}
\shorttitle{New insights into HD 5980}
\fulladdresses{
}

\listofauthors{M. Corcoran, T. H. Dall, C. Foellmi, A. Fullerton, L. Georgiev, A. Kaufer, 
G. Koenigsberger, S. Marchenko, D. Massa, P. Massey, A. Moffat, N. Morrell, Y. Naz\'e, J. Pittard, 
A. Pollock, N. St-Louis, O. Toledano  }
\indexauthor{Corcoran, M.}
\indexauthor{Dall, T.H.}
\indexauthor{Foellmi, C.}
\indexauthor{Fullerton, A.}
\indexauthor{Georgiev, L.}
\indexauthor{Kaufer, A.}
\indexauthor{Koenigsberger, G.}
\indexauthor{Marchenko, S.}
\indexauthor{Massa, D.}
\indexauthor{Massey, P.}
\indexauthor{Moffat, A.}
\indexauthor{Morrell, N.}
\indexauthor{Naz\'e, Y.}
\indexauthor{Pittard, J.}
\indexauthor{Pollock, A.}
\indexauthor{St-Louis, N.}
\indexauthor{Toledano, O.}

\SetVolume{00}
\SetYear{2008}

\ReceivedDate{2007 July 6 }
\AcceptedDate{}
\resumen{Presentamos los resultados de una campa\~na de observaci\'ones \'opticas del
sistema m\'ultiple HD 5980 cuya estrella primaria sufri\'o una erupci\'on importante
en 1994.  Atribuyendo la variabilidad de las l\'{\i}neas en emisi\'on al movimiento
orbital de las dos estrellas con \'orbita de 19.3 dias, deducimos sus masas.  El
comportamiento de las l\'{\i}neas fotosf\'ericas indica que la fuente de ``tercera luz"
del sistema es probablemente tambi\'en un sistema binario como fu\'e propuesto por
Scheickhardt (2002).  Los datos presentados en este art\'{\i}culo estar\'an publicamente
disponibles.}

\abstract{We present the results of optical wavelength  observations of the unusual SMC  
eclipsing binary system HD 5980 obtained in 1999 and 2004--2005.  Radial velocity curves for 
the  erupting LBV/WR  object ({\it star A}) and its close WR-like companion ({\it star B}) are 
obtained by deblending the variable emission-line profiles of NIV and NV lines.  The derived 
masses  $M_A=$58--79 M$_\odot$ and $M_B=$51--67 M$_\odot$, are more consistent with the the 
stars' location near the top of the HRD than previous estimates.  The presence of a  wind-wind 
interaction region  is inferred from the orbital phase-dependent behavior of He I P Cygni 
absorption components.  
The emission-line intensities  continued with the declining trend previously seen in UV spectra. 
The behavior of the photospheric absorption lines is consistent with the results of Schweickhardt 
(2002) who concludes that the third object in the combined spectrum, {\it star C}, is also a binary 
system with $P_{starC}\sim$96.5 days, $e=$0.83. 
}           
\addkeyword{Binaries: close}
\addkeyword{Stars: mass loss}
\addkeyword{Stars: individual (HD5980)}

\begin{document}
\RescaleTitleLengths{0.95}
\maketitle

\section{Introduction}

The Small Magellanic Cloud system HD 5980 consists of two very luminous  components 
referred to as  {\it star A}  and  {\it star B}  in a relatively close, eclipsing and eccentric 
orbit  and a third equally luminous source, referred to as {\it star C}, that may 
merely be a line-of-sight coincidence.  The eclipsing nature of HD 5980 was discovered by
Hoffmann, Stift \& Moffat (1978), although the actual 19.3-day orbital period was found by 
Breysacher \& Perrier (1980; henceforth BP80), and more recently refined by  Sterken \& Breysacher
(1997).  The two eclipses were not separated equally in time, indicating an eccentric  orbit 
($e=0.32$, Breysacher \& Perrier 1991; $e=0.27$ Moffat et al. 1998; $e=0.30$ Kaufer et al. 2002). The 
presence of strong and broad emission lines as well as  photospheric absorption lines led to its 
early classification as a WN+OB binary system (Azzopardi \& Vigneau 1975; Walborn 1977; 
Breysacher \& Westerlund  1978). However, Niemela (1988) showed that the Doppler shifts
in the N IV 4057 and N V 4603 emission lines moved in anti-phase, concluding that {\em both}
stars in the 19.3-day binary orbit were Wolf-Rayet stars of the nitrogen sequence (WN) and
that the absorption-line spectrum had to arise in a third object.  With this result, 
Niemela et al. (1997)  estimated  the masses of the two stars, M$_A\sim$50 M$_\odot$ and 
M$_B\sim$28 M$_\odot$.  From the eclipse light curve, Breysacher \& Perrier (1991, henceforth
BP91) derived stellar radii, R$_A\sim$21 R$_\odot$ and R$_B\sim$15 R$_\odot$. Its location in
the SMC implied a visual absolute magnitude M$_V^{sys}=-$7.3 mag (BP91) and the  BP91 light curve analysis 
provided good constraints on the  visual absolute magnitude for each of the three stars within 
the system, from which M$_V^A=$-6.3 mag, M$_V^B=$-5.8 mag, and M$_V^C=$-6.1 mag. 
These and other parameters for HD 5980 are summarized in the first column of Table 1, which
contains information derived from data obtained prior to  1981.  Results from data of other 
epochs is listed in column 2 (1994; near the time of eruption) and in column 3 (1999-2005;
declining phase of the eruption).

The spectral characteristics and visual brightness of HD 5980  gradually changed between  the late 1970's 
and 1993, when it entered an eruptive state that lasted $\sim$1 year (Bateson \& Jones 1994; 
Barb\'a et al. 1995; Koenigsberger et al. 1995).  The  activity  involved an increase in 
visual brightness and mass-loss rate, and a decrease in  wind velocity  and  effective 
temperature, similar to the eruptive phenomena observed in LBVs.  The radial velocity (RV) 
variations  in the  post-eruption emission-line spectrum  led Barb\'a et al. (1996, 1997)  to 
conclude that the instability producing the outbursts originated in {\it star A}. This was also  
confirmed by independent observations of Moffat et al. (1998).   The  hydrogen abundance 
in {\it star A} that was derived  near the time of maximum eruption, when its emission-lines are believed 
to have completely dominated the combined spectrum,  was found to be depleted  (Koenigsberger et al. 1998b). 
Thus, {\it star A} is a very massive star that has left  the Main Sequence.  Another fascinating aspect of 
the system is that in the process leading up to the eruption, {\it star A} seems to have traversed the entire 
nitrogen-sequence (WN) of the WR classification system, starting out as a WN3 (Niemela 1988) and 
reaching WN11 (Drissen et al. 2001) during maximum eruption.  This behavior is unprecedented.

It is important to note that the eruptive activity in HD 5980 has been recorded only once. Thus,
{\it star A} may  still be basically in the pre-LBV evolutionary phase, resembling one of the  
H-rich (or at least not yet H-poor), luminous WR stars of type WNh, or probably even WNha,
like most, if not all of the 10 other WN stars in the SMC (Foellmi et al. 2003).  These are 
not really {\it bona fide} WR stars in the classical sense, i.e. in the He-burning
stage. 

{\it Star B} is also inferred to be a WR star based on pre-eruption spectroscopic observations that 
indicated emission-line RV variations consistent with this notion (Breysacher, Moffat \& Niemela 1982).  
If both stars are, indeed, WR's, the time-variable wind of {\it star A} presents  a unique 
opportunity for studying the effects that a changing wind momentum ratio  produces on 
the wind-wind interaction (WWI).  Curiously, however, the change in the WWI region  during
declining phases of the eruptive event appears to have had little effect on the observed
X-ray emission (Naz\'e et al. 2007).

The masses of {\it star A} and {\it star B} are the central issue for establishing their 
evolutionary state.  But a vexing problem arises when radial velocity curves of the emission lines 
are used to derive the stellar masses.  The  centroid of an emission line is ill-defined 
when the line is asymmetrical and undergoes strong variability over the orbital cycle, as 
is the case for HD 5980.  Without  a clear understanding of the line profile variations (lpvs), 
the orbital motion cannot be reliably ascertained and, hence, one cannot establish the primary 
cause of the variability.

The objectives of this paper are to use new and archival observations to  establish constraints 
on the sources of the line-profile variability in HD 5980 and thus  obtain  RV curves that
may be more representative of the orbital motion.  Section 2 summarizes the  data; in Section 3
we describe the line profile variability, leading to a new set of RV curves (Section 4) from which the
masses of {\it star A} and {\it star B} are estimated;  Section 5 summarizes the evidence suggesting 
that the wind density of {\it star A} is diminishing over long timescales; Section 6 describes the 
stationary photospheric absorptions; Section 7 presents the results of the photometric monitoring; 
in Section 8 we discuss the wind-wind interaction region; and in Section 9 we present the conclusions.

\section{Observational data} 

The 2005/2006 observation campaign  at ESO  was coordinated by CF and GK. 
Spectroscopy was obtained under service observing mode with  the Fiber-Fed 
Extended Range Optical Spectrograph (FEROS)   mounted on the MPI 2.2m telescope  at 
La Silla,   and the ESO Multi Mode Instrument (EMMI) mounted on the NTT.    
In addition to these data, a set of archival FEROS observations obtained in 1999 
on the 1.5m telescopes at La Silla (Schweickhardt, 2000; Kaufer et al. 2002) were 
retrieved.  Photometric observations were obtained  at Las Campanas in 2003, at 
ESO in 2005, and at CTIO in 2005-2006. 

\subsection{Spectroscopy}

The 33 FEROS echelle spectra reported in this paper were obtained with a 
2 k $\times$ 4 k EEV CCD, and they cover the whole optical range ($\sim$3750-9200 \AA). 
The spectral resolving power of FEROS is 48000. Typical exposure times for the 1999 data were
between 40 and 50 minutes while those of the 2005 data were typically 27 minutes.  The 
S/N$\sim$30-70 per resolution element in the collapsed 1D spectrum, lower values corresponding 
to the blue spectral region. Tables 2 and 3 contain the ephemeris of these observations.

The ``pipeline" FEROS data reduction  was checked by LG against sample detailed processing using
MIDAS and IDL  routines and was found to provide similar results, and thus we chose to use
the ``pipeline" reduced spectra.  Data analysis was performed by GK.  The  merged spectrum 
in the wavelength range 3900--8500 \AA\  was  
normalized  in a three-step process.  In the first step, all the  spectra were combined 
for each epoch (1999 and 2005, respectively) so as to automatically remove cosmic rays.   
In the second step, the continuum was traced using a linear interpolation between
small wavelength intervals over which the location of the continuum could be relatively
well established.   This yielded a curve that approximately describes the shape of the
continuum on the average spectrum and that was then used to rectify the continuum on each 
of the individual spectra.  After the spectra were normalized in this manner, the third 
step consisted in fitting  a spline function (``spline3" in IRAF) of order 3, 6 or 15 to 
arrive at the normalized spectrum for the whole wavelength range stated above.  However, 
for the detailed comparison of the line profiles and  RV measurements, selected 
spectral windows were extracted from the whole normalized spectrum, and renormalized 
individually by fitting a second order Legendre function to the neighboring continuum levels.
Cosmic rays in spectral windows to be measured were removed individually, and a Gaussian smoothing 
procedure was applied to enhance the S/N ratio per wavelength interval.  

Spectra obtained near mid-continuum eclipses need to be renormalized 
so that the comparison of line profiles  is  meaningful.  Because the orbit is eccentric,
eclipses are not equidistant in phase; they occur at $\phi=$0 ({\it star A} ``in front") and
at $\phi=$0.36 ({\it star B} ``in front"). From the Breysacher \& Perrier (1991; henceforth, BP91) light curve
solution, $\phi=$0 eclipse is total while at $\phi=$0.36, only $\sim$50\% of {\it star A}'s continuum 
is occulted.  The relative continuum levels for stars A, B and C are, respectively, I$_A=$0.41, I$_B=$0.26
and I$_C=$0.33. Hence, to correct for the continuum eclipses, we added I$_B$ at $\phi=$0.00 and 
0.7I$_A$ at $\phi=$0.36--0.37 (see below), and then renormalized the continuum to unity.   
It must be emphasized that this renormalization is only a gross approximation, since no recent (i.e., for data
taken after 1980) light curve solution is available to provide current values of the stellar radii.  However,
under the assumption that only {\it star A} changed its size, it is pretty certain that the eclipse at $\phi=$0 is
a total eclipse.  In addition, FUSE observations lead to the conclusion that the ratio of {\it star A} and {\it star B}  
radii in 2002 is similar to that obtained by BP91.  This would suggest that the correction at this
phase be only 0.5I$_A$. However, this correction leads to an increase in the line-to-continuum ratio at
eclipse, contrary to what is expected, especially since a fraction of {\it star A}'s line emission must be
eclipsed, leading to the conclusion that a larger correction is required. This leads to the correction value
of 0.7I$_A$ given above.  Such a larger correction is justified when limb-darkening is taken into account.   


The ESO Multi Mode Instrument (EMMI)  was used in the red mode (RILD)  with the low-resolution grism 2, 
leading to a resolution of 300-1700 in the wavelength range 4000-10000 \AA. Typical exposures 
were  20s for HD5980. 
EMMI data were processed by OT using standard IRAF routines. This processing consists of cosmic 
ray excision, bias subtraction, flat-field correction, wavelength calibration,  background/sky  
subtraction and  extraction to produce a 1-dimensional spectrum.  Background includes the strong 
ISM lines that are produced by the large H II region N66 within which HD 5980 is located. Three 
spectra were obtained on each night, and these were averaged together and normalized using an 
interactive polynomial fit  to line-free continuum regions.  Wavelength calibration was done with the
HeAr lamp exposures obtained on the same nights as the exposures of HD 5980.
 Table 4 contains a summary of the EMMI data, and Figure 1 illustrates the  4000-7500 \AA\ region 
of the average EMMI spectrum at orbital phase 0.00.  Prominent emission lines are identified. 

\subsection{Photometry}

Observations  with the Danish 1.5m telescope, using DFOSC and the V Johnson filter, were 
coordinated by TD  and obtained over the time period JD 2453561 --2453693. DFOSC is a focal 
reducer imager and spectrograph, covering a 13 $\arcmin\times$13$\arcmin$ field of view. Data 
processing was carried out by SM as follows. The available CCD images were reduced with standard 
IRAF procedures. Following the general approach of Everett \& Howell (2001),  a weighted 
combination of fluxes of multiple comparison stars was introduced in order to produce differential 
magnitude values, $<$m$_v$-C$>$. 
This approach provides a typical accuracy $\sigma_{instr}$=0.003-0.008 mag for each 
individual observation,  as listed in column 5 of Table 6.
The additional columns in Table 6 list,  in columns 1-3 the average MJD of the observations, the orbital phase, and 
the number of differential magnitude determinations that were averaged to yield the $<m_v-C>$
listed in column 5. The standard deviation, s.d.,  for this average is listed in column 6. 

In addition to the above,   we include V-band  observations obtained by NM from 
JD 2452975.56 to 2452991.57 on the  1-m Swope Telescope, at Las Campanas Observatory, and
data that were taken from JD 2453591.88 through 2454092.70  for PM  by the SMARTS consortium on the 
CTIO/SMARTS 1.0-m telescope.  Both sets were processed by PM and converted to the visual  magnitude scale.
Part of the SMARTS data overlap in time with the Danish Telescope photometry, filling in
the gaps in orbital phase not covered by the latter.
The photometry of the Swope and SMARTS data were done relative to other stars
in the NGC 346 cluster, with the zero-point set by comparison of the data from a single
(arbitrary) exposure to the photometry of Massey et al.\ (1989).  Within  each data
set, the photometry has a relative accuracy of $\sim$0.02 mag. 

\section{Line profile variability}

Line-profile variability in WR binaries is a major obstacle for determining the stellar masses.
This is because  the amplitudes of the RV curves  obtained from  variable and asymmetric line
profiles depend on the way in which the lines are measured.  In HD 5980, emission line-profile 
variability  has been a common feature since the star was first observed spectroscopically 
(Breysacher \& Westerlund 1978).  Remarkably, the nature of the variability has remained 
qualitatively the same over many years, despite the strong changes that have occurred in the 
stellar wind properties.  For instance,   emission lines are always broader near 
elongations,  while  near eclipses they  are visibly narrower and more sharply peaked 
(BMN80; Moffat et al. 1998), as illustrated by  the Balmer-series lines of hydrogen that are 
plotted in Figure 2.  The left panels of Figure 2 correspond to eclipse phases and those 
on the right to orbital phases near elongations.  Most of the emission lines display the same 
pattern of variability. 

The change in full-width at half maximum (FWHM) is illustrated in Figure 3 
obtained from  the Gaussian function fits to HI (+He II) 4101, N IV 4057, He II 5411, and 
CIV 5806/12 \AA\ as a function of orbital phase, showing that different lines behave differently.  
However, the qualitative nature of the variations we observe in 1999 is as  reported in previous 
optical observations (Moffat et al. 1998; Breysacher et al. 1980; Breysacher 2000). This effect 
is only one aspect of a much more complex phase-dependent line profile variability pattern that 
involves  asymmetric distortions in the line shape.  Thus, the amplitude of any RV curve  derived 
from the emission lines depends on how the lines are measured; that is, it depends on  where within 
the system the emission lines are assumed to originate.

Fortunately, optical and UV spectroscopic monitoring of HD 5980 since the late 1970's provides
the first of three  pieces of solid information to aid in interpreting the variability:  numerous 
emission lines that are currently present in the spectrum were absent prior
to the early 1980's. The RV variations of the ``new" lines indicate that they arise in {\it star A}. 
This is also  true of the ``new" UV lines observed with IUE and HST (Koenigsberger 2004 and
references therein).  

A second fact is  that lines arising primarily in the inner portions of the stellar wind are less
susceptible to distortions due to  wind-wind interaction effects than  lines that arise from 
outer wind regions.   The former tend to originate from  atomic transitions between excited
states, and from elements of lower chemical abundance and are visible only because the
density of the inner, accelerating portions of the wind is sufficiently large.
Hence, lines from abundant ions and having large transition probabilities, such as H$\alpha +$He II 6560 \AA\
and He~II 4686 \AA\ should be avoided during the RV curve analyses, while lines such as  N~IV 4057 \AA\ 
may better describe the orbital motion.   

The third datum that aids in the interpretation of the lpv's is that the system is eclipsing and
the  largest RV's occur  near orbital phase $\phi=$0.15. In addition, due to the geometry of the orbit, 
during the eclipse at $\phi=$0.00 only the wind of {\it star A} contributes 
emission or absorption at  wavelengths shorter than the rest wavelength.  

Figure 4 provides a guide to the location of {\it star A} with respect to {\it star B} (located at the center 
of this coordinate system) as seen by an observer located at the bottom of the page.  The 
red squares that are labeled with  the orbital phases correspond to the location of {\it star A}
at that given time.   Periastron passage occurs near $\phi=$0.07.

\subsection{Emission lines}

With the above ideas in mind, we compared the 2005 FEROS line profiles obtained at 
orbital phases 0.13, 0.36 and 0.80  with the profiles observed at $\phi=$0.00.
The spectrum at $\phi=$0.0 is  expected to be dominated by {\it star A} emission.
The individual line profiles are illustrated in Figures 5--10,
plotted on a velocity scale corrected for a SMC motion of $+$150 km/s.  The
top left panel contains the $\phi=$0 spectrum.   The other three orbital phases
are presented in the remaining panels, and compared with the $\phi=$0 profile, which has
been shifted in velocity scale to align major features. For example, Figure 5 
displays the   He II 4542,  N V 4603 and N V 4621 \AA\ lines.  The velocity shift 
required to align the maxima of the three emission  lines at $\phi=$0.13  is $+$275 km/s,
while for  $\phi=$0.8, the He II 4542 \AA\ emissions are aligned with a shift of 
$-$150 km/s.   

We first consider the profiles of  N V 4603, 4621 (Figure 5) and the 
weak He II lines in the $\lambda\lambda$6000--6200 \AA\ region (Figure 6). At $\phi=$0.13 excess  
emission is apparent in the blue emission line wings, pointed out by arrows. 
In  N V 4603, 4621 \AA\ it can be seen as narrow sub-peaks that ``fill in" 
the P Cygni absorption components.  A similar excess emission  at this phase appears 
in  N IV 4057, H$\delta$ (Figure 7), H$\gamma$ (Figure 8), He II 5411
(Figure 9) and C IV 5806-11 (Figure 10).  We interpret this excess blue emission
as emission arising in {\it star B}, which is approaching the observer at $\phi=$0.13.

No relative shift was applied to the spectra at $\phi=$0.37 since the N IV 4057
line coincides almost exactly with the profile of $\phi=$0. In most of the other
lines, at least the red wings match very well, but blueward of line center one can
easily see that excess emission is present.

At $\phi=$0.8 the lines present two different patterns. In the first, N IV 4057, 
the weak He II 6000-6200 \AA\ emissions and the N V 4603 P Cygni absorptions  match 
the shifted $\phi=$0 line profiles.  Thus, these features are primarily formed in 
{\it star A} and have negative RV's due to its approaching orbital motion.
The second pattern, however, is more prevalent and consists of the presence of 
excess emissions on both wings, although the excess red emission is generally
more extended than the blue.  Within the context of the interpretation given above, 
the excess red emission at $\phi=$0.8 originates in {\it star B}, which is receding from the 
observer at this phase.  The excess blue emission, however,  appears 
to arise from an additional source.  It is tempting to suggest that it 
arises from the same source as that which produces excess emission at $\phi=$0.36, although 
its projected velocity towards the observer is greater at $\phi=$0.8 than at  $\phi=$0.36.

Figures 11--14 are grey-scale representation of the 1999 FEROS data and provide a global view 
of the lpv's in 4 representative lines: N IV 4057, He II 5411, H$\beta$+He II 4860 and He I 6678 \AA\.
They are constructed with  line profile residuals, stacked from bottom to top with increasing 
orbital phase.  The residuals are the difference between each individual spectrum and the average 
of the  15  spectra obtained over a single orbital cycle and listed in Table 2 (the first two 
spectra of this table were not included here). The spacing between the spectra  
is $\Delta\phi\sim$0.05 in orbital phase, and to consistently retain this spacing in the
dynamic plots,  the gaps of missing phase coverage are filled-in with grey 
background color. Plotting the same data twice provides a clearer picture of  the orbital-phase 
dependent trends.  The white regions in these images correspond to excess emission; the black 
to emission deficiency with respect to the mean, or absorption. The maximum dynamic range is 
typically 10--15\% in the residuals with respect to the normalized mean profiles.

All four dynamic plots display similar large-scale trends, specifically, the excess emission  
traces out a  curve suggestive of orbital motion associated with {\it star A}.  Some of the
details of the plots, however, are different.  For example, there is a fainter  sinusoidal
curve that moves in anti-phase with the more dominant bright sinusoid that is  clearly seen in
He I 6678 and H$\beta$+He II 4860, weaker in He II 5411, but not detectable in N IV 4057. In 
He I 6678, this secondary curve has a "whisp-like" structure around $\phi=$0 in the velocity
range 0 -- 1000 km/s.  A striking difference between N IV 4057 and H I 6678 is the very extended
blue emission in the phase interval $\sim$0.9--1.0 in the latter.  While N IV 4057 shows a rapid
shift towards the red, as expected from the orbital motion of {\it star A}, He I 6678 stays very much
the same, extending over the velocity range $-$300 to $-$1100 km/s.  This implies the presence
of emitting gas that is approaching the observer with a speed that is not correlated with
the orbital motion within this phase interval.


\subsection{ Peculiar behavior of He I P Cygni absorptions}

A clue to the origin of the excess blue emission at $\phi=$0.80 may lie in the
behavior of the He I lines.  As an example, consider the blue wing of He I 5875 \AA\ 
that is shown in Figure 10. At $\phi=$0.80, a sharp P Cygni absorption is present in 
the same velocity range as the excess blue emission in the H I and He II lines shown 
in the grey-scale plots discussed above. Such sharp and strong P Cygni absorptions are 
usually associated with lines that are formed far from the star, in the low-density 
regions that are expanding at a constant speed, usually  the terminal wind speed, v$_\infty$.  
The presence of the absorption would not be surprising were it not for its {\em absence} or 
weakness at other orbital phases.  This effect can be more clearly appreciated in Figures 15, 
which displays montages of He I 4471, 5875, 6678 and 7965 \AA\ at the four orbital phases 
$\phi=$0.0, 0.13, 0.37 and 0.80, and in Figures 16 and 17 where the fuller orbital phase
coverage of the 1999 data shows that the absorptions are strongest in the orbital phase
interval 0.7--0.9.  They nearly vanish at $\phi\sim$0--0.10 (around periastron passage),  
and re-appear briefly right before the secondary eclipse at $\phi=$0.36.  They are  then 
very weak or absent (or greatly displaced) between $\phi=$0.36 and $\sim$0.7. 

There is no immediate explanation for this behavior, although similar features as described 
here have been linked to the WWI interface in the well-documented case of V444 Cygni  
(Marchenko et al. 1994, 1997),  GP Cep (Demers et al. 2002),  $\gamma^2$ Vel (De Marco, O., 2002), 
and $\eta$ Carinae (Nielsen et al. 2007). It is also interesting to note that in these four 
systems the observations lead to the conclusion that there are substantial differences in 
strengths and velocities of the leading/trailing WWC arms.  A similar scenario for HD 5980 
will be discussed in Section 8.

In summary, we find that the only lines in the optical spectrum of HD 5980 that are likely 
to  truly describe the orbital motion of {\it star A}  are N IV 4057, N V 4603 and
some of the weak He II lines in the $\lambda\lambda$6000--6200 \AA\ range.   
 
\section{New constraints on the masses of the system}

The top panel of Figure 18 illustrates the RV curves that result when the centroid
of the emission lines is measured using a single Gaussian fit.  The corresponding data
are listed in columns 4, 7 and 10 of Tables 2 and 3.  The RV curve for N IV 4057 \AA\ line 
is very similar to that derived by Barb\'a et al. (1997), indicating that {\it star A} is 
still the major contributor to this line.  The H I $+$ He II 4101 and He II 5411 \AA\ display a 
similar trend, but with different  amplitudes and shapes, implying that they contain a more
significant contribution from {\it star B} and from the WWI region.
     
Assuming that the line profile variability is in part due to the superposition 
of emission  arising in  two distinct sources, we deblended the lines using a two-function fit to 
the  profiles.  
The results of these measurements are plotted in the middle and bottom panels of Figure 18 
(filled-in pentagons), showing two RV curves in anti-phase.  Using a similar procedure, we 
deblended the N V 4603 emission (open stars) using two Gaussians. The results for N V are more  
uncertain for two reasons: 1) the P Cygni absorption component of the neighboring N V 4621 \AA\ 
line limits the extent of the N V 4603 \AA\ red wing; 2)  at $\phi=$0.13, {\it star B}'s component 
is shifted into {\it star A}'s P Cygni absorption component. On some spectra, this superposition 
makes it very difficult to measure the secondary component, since it simply fills in the P Cyg 
absorption but is not prominent enough to enter the two-function fit. Because {\it star A}'s 
contribution is dominant, its RV curve is not severely affected by these two effects.  However, 
that of {\it star B} not only displays much larger scatter, but also a systematic deviation from
the RV curve derived from N IV 4057.  Part of this deviation may also be due to NV emission
arising in the WWI region.

The middle panel of Figure 18  also contains the RV variations of  the weak  He II 6074.1 and 
He II 6118.2 \AA\ lines (small open squares), obtained by fitting a single Gaussian function.  
These lines show a slightly smaller RV amplitude than that of N IV 4057 \AA\,  consistent with 
the notion that they arise primarily in the wind of {\it star A} but also contain a contribution 
from {\it star B}.  

The shape of the RV curve of {\it star B} shown in the bottom panel of Figure 18  matches quite 
well the mirror of the RV curve for {\it star A}, within a constant factor (related to the relative 
masses of the 2 stars), thus supporting the hypothesis that the excess emission (with respect to 
the $\phi=$0 spectrum) is in fact primarily coming from the 2 stars.  The WWI excess should be  
out of phase with the orbital motion.

We performed a fit to the RV amplitude of NIV and NV using the genetic algorithm PIKAIA 
(Charbonneau, 1995).  The radial velocity curves of the two stars were fitted simultaneously, 
giving the two RV semi-amplitudes (K1 and K2), the eccentricity $e$, and the longitude of the 
periastron $\omega_{per}$. 
The  semi-amplitudes are K$_A=$198 km/s (NIV) and 180 km/s (NV), and K$_B=$222 km/s from both 
lines.  These results lead to $M_A=$58--79 M$_\odot$ and $M_B=$51--67 M$_\odot$, adopting an 
orbital inclination angle  $i=$88$^\circ$ (Moffat et al. 1998), and given the uncertainties
in the fit to the RV curves ($\pm$4 km/s and $\pm$16 km/s for {\it star A} and {\it star B},
respectively).  The corresponding range in semimajor  axis is $a=$143--157 R$_\odot$.  The 
orbital eccentricity derived from both N IV and NV solutions is $e=$0.30$\pm$0.16.  This value 
is compatible with $e=$0.27 (Moffat et al. 1998) and $e=$0.32 (PB91).  The longitude of periastron 
that we derive lies in the range $\omega_{per}=$283$^\circ$--294$^\circ$.  This is  smaller than 
the corresponding values of $\omega_{per}=$319$^\circ \pm$6 (Moffat et al. 1998) and 
$\omega_{per}=$313$^\circ$ (BP91). Given that the RV curve has a limited orbital phase coverage 
and that various interaction effects are present in the emission-line spectrum, we favor the 
values of $\omega_{per}$ derived from the photometric light curve and polarimetry (BP91 and 
Moffat et al. 1998).

The above masses  are significantly larger than  those   based on single function fits to the
emission lines and they are more consistent with the stars'  position on the H-R Diagram near the
120 M$_\odot$ evolutionary track (Koenigsberger 2004).

\section{Declining wind densities ?}

The emission lines in 2005 are substantially weaker than observed in 1999, indicating that
the trend for decreasing wind densities that was observed in UV spectra (Koenigsberger 2004)
is continuing.   In order to quantify the effect, we measured the equivalent width (EW) of
H $\delta +$HeII 4101 \AA\, NIV 4057 \AA\, and He II 5411 \AA\ as listed in Tables 2 and 3. 
The orbital phase-dependent variability in 1999 is plotted in Figure 19, showing that 
He II 5411 \AA\ displays a large modulation  with a broad minimum centered around $\phi\sim$0.15, 
while  N IV 4057 \AA\ gives only a hint of an eclipse at $\phi=$0.36.  No clear phase-dependent 
changes are evident in H$\delta +$He II 4101 \AA\.   
Figure 20 illustrates  the ratio  EW(He II 5411)/EW(H$\delta +$He II 4101) as a function of
orbital phase for the two different epochs, 1999 and 2005, leading to the conclusion
that the degree of ionization in HD 5980 is greater in 2005 than in 1999.
Consistent with this conclusion is the significanly weaker N III 4640 \AA\ blend in 
2005 compared to that of 1999. 

Hence, these are indications that HD 5980's wind is  reverting to its state prior to the 1993/1994 
eruptions, when the emission lines were much broader and significantly weaker (Koenigsberger
et al. 1998a; Moffat et al. 1998). 

\section{Stationary photospheric absorptions}

A striking feature in the 2005 FEROS data is the strength of  photospheric absorptions 
clearly seen superposed on the  emission lines  (see, for example, Figures 2, 7,
8, 10, and 15).  
Similar absorptions have been observed in 1999 HST/STIS spectra (Koenigsberger et al. 2003), 
and  found to be relatively stable in wavelength; that is, the amplitude of the RV variations was 
$\leq$40 km/s, which was  within the uncertainties of the RV measurements.
Table 5 lists the dominant absorption lines that we identified in the 2005 FEROS data, with
measurements of their RV (columns 2,4,6, and 8) and their corresponding FWHM (columns 3, 5, 7, and 9),
measured with Gaussian fits to the profiles.  It is important to note that since these features are 
superposed on strongly variable emission lines,  the extent of the absorption line wings is nearly 
impossible to establish.  Hence, the RV measurements refer  to the line cores and these do not display
any significant RV variations.  In fact, if we average the RV values for each orbital phase
listed in Table 5, we find that the average RV values are virtually constant, within the standard
deviation of the measurements at each orbital phase.  The averages and standard deviations are listed
in the bottom rows of Table 5.  The average velocity at each of the 4 orbital phases listed in
this table are within the range 161--172 km/s, with a corresponding range in standard deviations
of $\pm$16 -- $\pm$22 km/s.  Hence, they are indeed stationary within the timescale of the 2005
observations.   

The photospheric absorptions are believed to arise in  {\it star C} (Niemela, 1988; 
Koenigsberger et al. 2001) which is a reasonable conclusion, given their apparently stationary 
nature over the P=19.2654 day orbital timescales.  Over longer timescales, however, the RV and 
strength of the lines  undergo changes, as illustrated in  Figure 21, where the He I 4471 \AA\ 
lines observed in 1999 and 2005 are compared.  At each of the four orbital phases plotted in the 
figure, the 1999 profiles are significantly redshifted   with respect to those of 2005.  Schweickhardt (2000) 
measured the RVs of the O III 5592.4 photospheric absorption line in  1999  FEROS data,
 finding them to be periodic with P$_{starC} \sim$96.5 days, and suggested that 
{\it star C} is itself a binary system.  His analysis of the RV curve implies that the {\it star C} 
system is very eccentric ($e=$0.83; $\omega=$255$^\circ$; T$_0=$2451183.6) and that the amplitude 
of the star responsible for the O III photospheric absorption is $K=$71 km/s. Similar results 
were derived from the analysis of the He I 4471 \AA\ photospheric absorption.  Our 2005 data 
only cover a fraction of the phase  for such a period, $\phi_{starC}\sim$0.4--0.8,
which lies in a relatively constant portion of the {\it star C} RV curve.  However,  OIII 5592.4  
RV variations do show a declining trend with total amplitude of $\sim$20 km/s which is 
consistent with Schweickhardt's (2000) conclusion. If this binary is gravitationally bound to the 
A$+$B pair or whether it is merely a line-of-sight projection remains to be resolved.  In any
case, however, Schweickhardt's (2000) results suggest that {\it star C} is also an intriguing
object that requires further analysis.

An explanation for  why the photospheric absorption lines  appear to be more prominent in
2005 compared with 1999 may lie in the decreasing  visual brightness of {\it star A} 
which thereby makes  {\it star C}'s photospheric absorptions appear more prominent. 
Another possibility is that since the emission lines became weaker in 2005 compared 
with 1999,  photospheric absorptions from {\it star A} could enhance the strength of 
absorptions in the combined spectrum.

In Table 5 we have also listed the measurements of a photospheric-like dip observed at
orbital phase $\phi=$0.13  ``blue-ward" of the stationary absorptions.  It is typically
located $-$500 km/s from the stationary absorptions, with a FWHM$\sim$500 km/s.  At this
phase, {\it star B} is approaching the observer and it is conceivable that these broad
features could be associated with its photospheric absorptions.  The large width and the
excess blue-shift (with respect to the orbital motion) could be produced by the
velocity gradient in its photosphere.   An alternative explanation may reside in the presence 
of a slower-moving wind region (or WWI region ?) projected upon the stars at this phase.
However, this issue must await further observations since, as the intensity of the emission 
lines from {\it star A} diminishes, the characteristics of photospheric absorptions will 
become easier to establish.  

\section{Light curve}
   
Figure 22 illustrates four sets of  HD 5980 photometry.  The best full orbital
phase coverage (open squares) consists of visual magnitude observations obtained using the 
CTIO/SMARTS  telescope during  2005.  The second set having full phase coverage  consists 
of the original Breysacher \& Perrier (1980) differential magnitudes (crosses), with a constant value of 
10.415 mag added in to make the average levels coincide with those of the SMARTS and the Swope data at orbital 
phase 0.60.  The Swope data (small plus signs) has a poorer overall phase coverage, but covers  
portions of the ascending branch of the $\phi=$0.36 eclipse more densely.  Finally, the Danish telescope
differential magnitudes to which the constant value of 12.57 mag has been added (triangles)  are primarily 
concentrated around $\phi=$0.99 and around the $\phi=$0.36 eclipse phase. The top
panel of Figure 22 illustrates the complete light curve, while the middle and lower panels focus
on the two eclipses.

The difference between the 2003 (SWOPE) and 2005 SMARTS data sets during the ascending
branches of both eclipses (there is practically no 2003 data during the descending branches) is
noteworthy.  It shows the large degree of epoch-to-epoch variability in the system. Also,
we should note that in order to shift the photometric minimum  to the anticipated  $\phi =0.36$,
a slightly corrected $T_{modified}=$2443158.865 is suggested, which also better accomodates
the photometry obtained in 1994-1995 (Moffat et al. 1998).   This  implies a significantly larger
difference with respect to the Sterken \& Breysacher (1997) T$_0=$2443158.707$\pm$0.07 than their
quoted uncertainty.  The possibility that this shift is related to apsidal motion and a full
analysis of the eclipse light curve will be explored in a forthcoming investigation.  It
is important to note   that the analysis requires a methodology as that used by Breysacher \& Perrier 
(1991) in which the extended nature of the continuum eclipsing sources is contemplated.

The longest train of Danish  data was obtained around eclipse at $\phi=$0.36--0.37, and the individual
data points are plotted in Figure 23, disclosing  $\sim$0.$^{mag}$035 amplitude variations  within a 
timespan of half an hour.  This is considerably shorter than the $\sim$6-hour periodic variations 
(0.02 mag) reported by Sterken \& Breysacher (1997) and attributed to  oscillations of {\it star A}.   
Curiously, if we assume T$_0=$T$_{modified}=$JD 2443158.865, then the two prominent minima appear symmetrically
located around $\phi_{modified}=$0.36 (where $\phi_{modified}$ is the  phase computed using the modified
value of T$_0$).  This would seem to indicate the presence of a bright emitting source that is
eclipsed just prior to and just after mid-eclipse, but not during mid-eclipse.  Or, alternatively,
the presence of dense, eclipsing gas located  at a distance from {\it star B} that is slightly larger
than the radius of {\it star A}.   One geometry for such a gas that comes to mind is that of a WWI
interacting region whose apex is transparent to optical radiation.

Strong variability near $\phi=$0.36 on similar timescales has previously been reported in polarimetric 
observations (Villar-Sbaffi et al. 2003), and two HST/STIS UV observations obtained one day apart
at these same phases suggest that the wind speeds are very variable. 

\section{The wind-wind interaction region}

The current view of the HD 5980 system, consisting of two stars with powerful winds, implies
the presence of an interaction zone.  Thus, at least part of the 
variability reported in this paper may result from phase-locked wind-wind interaction effects.  
In its simplest representation, the WWI region can be modeled in terms of a discontinuity
between the two winds that consists of the two thin shocks formed when the supersonic winds
collide and whose shape depends on the momentum ratio of the stellar winds.  Figure 24
is a cartoon illustrating the WWI region shape for HD 5980, assuming that the wind of {\it star A}
dominates. This assumption is based on the fact that all emission lines in the combined spectrum
increased significantly in strength  after the 1994 eruption and that the phase of their Doppler motions  
corresponds to that of {\it star A} (Koenigsberger 2004). The significantly larger emission measure 
from {\it star A} compared to {\it star B} is expected to correlate with a larger density.
Preliminary stellar wind models for the HD 5980 system that have been fit to  FUSE 2002 data 
imply that {\it star A}'s mass-loss rate was still large, $\dot{M}_A\sim$10$^{-4}$ M$_\odot$/yr 
(Georgiev et al., in preparation).   In addition, for {\it star B}, Moffat et al. (1998) found
$\dot{M}_B=$2$\times$10$^{-5}$ M$_\odot$/yr from polarimetric data obtained prior to the eruption.
These mass-loss rates, combined with wind terminal speeds derived from UV and FUV observations
lead to the WWI shock cone configuration shown in Figure 24, with a shock cone opening angle
$\sim$50$^\circ$.

 Four orbital phases are shown.  {\it Star A} is represented with three concentric
circles depicting, respectively, its photosphere (R$_{A}$), the extent of its accelerating wind region in
the case of a ``fast"  velocity law ($\sim$1.5 R$_{A}$), and the extent of the dominant  emitting
wind region for a moderately strong, but arbitrary, emission line ($\sim$4 R$_{A}$).   Note that 
because the orbit is eccentric, the WWI discontinuity intersects the line emitting region around 
periastron passage, but not around apastron.  When the WWI region intersects the line-emitting
region, the intensity of the observed line-profile is truncated at the corresponding velocity range.
The sets of dashed lines drawn parallel to the line-of-sight in Figure 24 enclose columns of
wind material that are projected against the luminous stellar continua and that are expanding
towards the observer (located at the bottom of the figure). All P Cygni absorption components observed in
the spectrum  arise in these columns.

The behavior of the blue-shifted, sharp absorption components of the He I emission lines 
suggests that they arise in the WWI region.  They are strongest during the orbital 
phase interval 0.7--0.9 (when both stars are in view and one of the WWI arms lies in the
foreground of {\it star B}) and they are  weak at $\phi\sim$0--0.10 (when mostly {\it star A} is in view)
and  $\phi=$0.36 -- 0.7 (when mostly {\it star B} is in view).  
It is important to note that the sharp and deepest portion of the P Cygni absorption components  
generally   arise in the  wind regions  that have attained terminal speeds.  In the 1999 data, at 
orbital phase $\sim$0.8, the He I 4471 \AA\ sharp feature extends between -900 to -1340 km/s, and 
maximum absorption at $\sim -$1130 km/s (assuming a correction for SMC motion of $-$150 km/s). This is 
significantly slower than terminal wind speed of $-$1750 km/s  inferred for {\it star A} during 
the same epoch from HST/STIS observations of UV lines (Koenigsberger et al. 2000).  This slower
speed is qualitatively consistent with absorption of {\it star B} radiation traversing the
foreground WWI region arm.  Also noteworthy is the fact that excess blue-shifted emission appears 
in many lines around orbital phase $\phi=$0.36, which is tempting to attribute to emission arising 
in the WWI zone that is not projected against either continuum source. 

Figure 25 provides estimates of the flow velocities, V$_t$, along the shock cone walls   derived 
from the expressions given by Cant\'o et al. (1996), using the roughly estimated values for mass-loss 
rates of the two stars.  At orbital phases around 0.8, the line-of-sight from the observer to
{\it star B} crosses  regions of the WWI having  V$_t \sim$600--800 km/s. Considering that the 
shock cone has an opening angle $\sim$50$^\circ$, the maximum projected speed of V$_t$ along the 
line-of-sight  in the stationary frame of reference is $\sim-$600 km/s.  This is significantly smaller
than the speeds at which the He I P Cygni absorption components are being observed.  Correcting for  the 
orbital motion of {\it star B} of $\sim$220 km/s brings the predicted value closer to the
observations at $\phi \sim$0.15, when {\it star B} is approaching the observer, but increases the
discrepancy round $\phi\sim$0.8.   However,     the idealized WWI  model presented 
in Figures 24 and 25 is a greatly simplified model. For example, if  we take into account that the 
axis of symmetry for the WWI shock cone is in reality tilted  with respect to the line connecting the 
two stars,  the angle between the leading WWI arm and the line-of-sight decreases, allowing for
faster projected speeds.  In addition, numerical simulations that include orbital motion show that
the shock cone arms are curved.   Hence, the leading and the trailing arms could cross the 
line-of-sight to the stars at at different orbital phases and with larger projected speeds than
depicted in Figure 25.  Figure 26 is a cartoon that illustrates an alternative geometry for the 
WWI region, based on an interpretation of the variable He I P Cygni absorption lines.  It would 
be interesting to analyze whether the physical mechanisms involved in the WWI can give rise to
such a different geometry, compared to the typical conical surfaces that are predicted by  the 
traditional calculations.

It is  important to  keep in mind that the geometry and characteristics of the interaction region 
depend on whether the shocked gas is radiative or not, and this, in turn,  determines the intensity 
and shape of line emission/absorption that the WWI region can produce.  One can estimate whether the 
gas is radiative or not using the parameter $\chi$ from Stevens et al. (1992), which is the ratio of 
the postshock cooling timescale to the flow time out of the system. Hydrodynamical models have indeed 
confirmed that the postshock gas cools rapidly when $\chi\sim$1.0.

To evaluate $\chi$ we need to determine the pre-shock velocity. $\chi$ is highly
dependent on this value ($\propto v^4$) so small changes in $v$ make a big
change in $\chi$. Using the approximation to the velocity law as in  St.Louis et al. (2005),

\begin{equation}
v = v_0 + (v_\infty - v_0)*(1 - R_*/r)^\beta
\end{equation}
\noindent with $v_0 = 0.16 v_\infty$ and $\beta=$2.5.

With such a velocity profile the wind accelerates so slowly that the
post-shock gas (from both stars) is always highly radiative, and the Canto et al. (1996) 
prescription used in Figures 24 and 25 is appropriate. However, if the wind(s)
accelerate quicker (eg $\beta=$1) then there is a chance that, for example, {\it star B}'s
shocked wind is adiabatic near apastron (but strongly cooling at periastron). 
%
%
%

The fact that the X-ray emission is apparently harder near phase
0.36 (Naz\'e et al. 2007) is very suggestive that the maximum pre-shock
velocity increases as the stars separate, although it is by no means certain 
that the winds between the stars bear much relation to the winds from single 
stars.  Indeed, indications exist that at least {\it star A}'s wind velocity 
structure facing the companion differs from that of its opposite hemisphere 
(Koenigsberger et al. 2006).

Unfortunately there have been very few hydrodynamic calculations of the wind
interaction in short period systems like HD 5980 where the acceleration of
the winds is taken into account (Iota Ori, Pittard 1998; Sand 1, St-Louis et
al. 2005). Antokhin et al. (2004) took a slightly different approach which
enabled them to  calculate the X-ray emission, but the wind needed to be 
specified by a beta velocity law and was not calculated self-consistently 
with inhibition and braking in the way that the other hydro models noted were.

If both winds rapidly cool after being shocked, the postshock
region collapses into a dense sheet and the shocks are coincident with the
contact discontinuity. It is worth noting  that in this case one expects
very strong instabilities in the WWI region (specifically the non-linear
thin-shell instability - see Stevens et al 1992), which means that the
position of the contact discontinuity from the Canto et al.  model should be
viewed only in a time-averaged sense. This is of course a possible cause of
the short-timescale fluctuations that are  mentioned in Section 7).

Wind clumping is an additional factor that is absent from the simplified view 
of WWI regions.  Pittard (2007) presents simulation of the effects of clumps in
binary systems with adiabatic colliding winds. Although the clumps are likely 
destroyed quite rapidly in such systems, they create a broader distribution of 
temperatures within the WWI at any given off-axis distance. This is because 
the interclump material is heated to higher than average temperatures, while the 
clumps are initially heated to significantly lower temperatures (the shocks 
driven into them are slower) - the temperature within shocked clumps is lower than the 
temperature obtained from the collision of smooth winds by a factor equal to the 
density contrast of the clump to interclump material. If most of the wind mass is within 
the clumps this should have observable consequences. The shocked material within the 
clumps is then heated further by additional weaker shocks and mixing within the WWI. 
In systems where the wind parameters are such that a WWI formed by smooth winds is 
not far from being radiative, the presence of higher densities within clumps may be 
enough for cooling to become important (see Walder \& Folini 2002).

\section{Conclusions}

In this paper we  present  results from  spectroscopic and photometric observations of 
HD 5980 obtained in  1999 and 2003-2005.  We first analyze the line-profile variability 
in the 2005 spectra, concluding that in lines that are formed in the dense, inner 
wind regions, line-profile-variability may be interpreted mostly in terms of 
the presence of two emission components, the first belonging to {\it star A} (the eruptor) and the 
second associated with {\it star B} (the companion).   Assuming that the emission-line profiles can be 
effectively separated with a two-function fit,  we find that the centroids of the two functions are 
consistent with the orbital motion of the two stars, leading to a new estimate of the masses,  
M$_A\sim$58--79 M$_\odot$ and M$_B\sim$51--67 M$_\odot$. These values are significantly larger than 
previously derived (Niemela et al. 1997), but still imply that the stars must have lost considerable 
amounts of mass before reaching their current evolutionary state, given their location on the HRD 
near 120 M$_\odot$ ZAMS evolutionary tracks (Koenigsberger 2004).  The mechanisms by which such 
large amounts of mass may have been stripped require detailed examination.  In addition, it would 
be very interesting to get more theoretical insight into their current internal stellar structure 
and hence, be able to establish their current evolutionary state.  As pointed out by Koenigsberger 
et al. (1998b), if {\it star B} is a {\it bona fide} WNE remnant \footnote{that is, the bare  He-
or C-burning core of a massive progenitor that has been stripped of its outer layers, as opposed 
to a massive star in an earlier evolutionary stage undergoing strong wind mass-loss}  of the 
originally more massive member of the binary,  its  interior  may have already reached a very 
advanced stage of evolution and, thus, it may be  nearing the supernova stage.  On the other hand, 
the significantly larger mass that we have now derived for {\it star B} (M$_B\sim$56--67 M$_\odot$ 
{\it vs.} $\sim$28 M$_\odot$ as believed previously) suggests that it may not  yet have reached the  
core He burning  evolutionary phase.  Since {\it star B} is believed to be the source of  the 
highly He-enriched WR-spectrum observed in the 1970's (Breysacher, Moffat \& Niemela 1982), the
implications for stellar evolution theory of such a massive, He-rich object are potentially
very interesting.

Line strengths in the 2005 spectra are significantly weaker than those
observed in 1999, consistent with a similar decreasing trend observed in UV lines, and
suggesting that the eruptor is reverting to a more quiescent state.   The question of 
whether another activity cycle will start in the near future is an important one,
since the timescale for the eruptive events provides an estimate of the total amount
of mass that may be lost through this mechanism.  In addition, the activity timescale
provides a clue to understanding the underlying mechanism responsible for the
instability.  Continued observations are required to determine  the long-term
variability pattern.

Visual photometric observations in 2005 provide a light curve covering the entire orbital 
cycle, as well as a detailed description of the system's brightness variations during 
selected orbital phases. In particular,  relatively large-amplitude ($\sim$0.035 mag) 
fluctuations appear around secondary minimum, when {\it star B}, the non-erupting star, 
is ``in front".  Continued systematic observational monitoring around this orbital phase would 
be very valuable because of the  strong and rapid variations  observed in visual 
photometry, polarimetry and UV P Cygni line profiles. In addition, the X-ray appears to
be greater here than around  the opposite eclipse.

The temporal behavior of the P Cygni absorption components of the  He I lines is qualitatively 
consistent with the presence of a denser wind structure folding around {\it star B} which we associate
with the WWI region.  However, the observed speeds are not consistent with the simple models.
In particular, the flow speeds are significantly larger than predicted in the radiative shock
limit and thus suggest that at least one of the shocks is adiabatic.  It is interesting to note 
that the hemisphere of {\it star A} that faces the companion is predicted to
be particularly active due to tidal interaction effects (Koenigsberger \& Moreno 2007),
raising the question of how this activity may affect the structure of the WWI region. 

The photospheric absorption lines remain stationary over the 19-day orbital  cycle.  However, we find a 
systematic shift of $\sim$20 km/s in  their radial velocities observed in 1999 with respect to those observed 
in 2005. This is consistent with the conclusion of Schweickhardt (2000) that {\it star C}, the third
source in the combined HD 5980 spectrum, is itself a binary system, with P$_{starC} \sim$96 days.
It is unclear, however,  whether the {\it star C} system is gravitationally bound to the 
{\it star A+star B} pair with a much longer orbital period.  It is  a curious fact that the {\it star C}  
orbital period is nearly a factor of 5 longer than the orbital period in the {\it star A+ star B} pair, which
may indicate that the two systems are physically bound.  If this were the case, then the eruptive
behavior in HD 5980 could be linked to the gravitational perturbations produced during periastron
passages. In this case the orbital period of the {\it star C} system around the {\it star A+ star B}
system would be $>$25 years, since only one large eruption has been seen over the $\sim$40 years
that the system has been studied.

In the Appendix we list the SWOPE and SMARTS photometry averages for each day of observation (Tables 7--10).
The data presented in this paper will be made publicly available (please contact CF or GK).

\acknowledgements

We express our gratitute to  Catherine Cesarsky  for having granted  European Southern Observatory 
Director Discretionary  observing time that provided the 2005 FEROS and EMMI observations  and
to the ESO observing staff for carrying out these  observations;  to Jens Hjorth for granting
time on the Danish telescope and Uffe Graae J\-orgensen, Daniel Kubas, Chlo\'e Feron and Christina 
Th\"one for performing the observations.  We thank Felix Mirabel for hosting the visit of OT to ESO in Chile. 
Some of the photometric data were obtained via the SMARTS consortium, and we are grateful to the various 
observers who contributed to this effort.  We acknowledge support to CF from the Swiss National Science  
Foundation; to GK from PAPIIT/DGAPA/UNAM IN 119205; to PM from NSF grant AST-056577; to AFJM and NSL from 
NSERC (Canada) and FQRNT (Quebec);  and to YN from the FNRS and the Prodex XMM-Integral contract (Belspo). 
This research was supported in part by the Danish Natural Science Research Council through its centre
for Ground-Based Observational Astronomy, IJAF/IDA, and   by the Gemini Observatory, 
which is operated by the Association of Universities for Research in Astronomy, Inc., on behalf 
of the international Gemini partnership of Argentina, Australia, Brazil, Canada, Chile, the 
United Kingdom, and the United States of America.

\vspace*{7pt}
\begin{itemize}

\item   Antokhin, I.I., Owocki, S.P., Brown, J.C. 2004, ApJ, 611, 434.
\item   Auer, L.H. \& Koenigsberger, G. 1994, ApJ, 436, 859
\item   Barb\'a R.H., Morrell, N.I., Niemela, V.S., Bosch, G.L., Gonz\'alez, J.F.,  Lapasset, E., Ferrer, O.E., Brandi, E.E., Cellone, S.A., Garcia, B.E., Malaroda,S.M.,  Levato, O.H., Donzelli, C., Feingstein, C., Rich, M. 1996, RMAACS, 5, 85.
\item   Barb\'a R., Niemela V.S., Morrell, N. 1997, in ASP Conf. Ser. 120,  Luminous Blue Variables:Massive stars in transition, ed. A. Nota \& H. Lamers  (San Frnacisco:ASP), 238.
\item   Bateson, F.M. \& Jones, A. 1994, Publ. Var. Star. Sec. R. Astron. Soc. New Zealand, 19.
\item   Breysacher, J.,François P. 2000, A\&A, 361, 231.  
\item   Breysacher, J. \& Perrier, C. 1980, A\&A 90, 207  (BP80)
\item   Breysacher, J. \& Perrier, C. 1991, in IAU Symp. 143,  143, Wolf Rayet Stars and Interrelations with other  Massive Stars in Galaxies,  eds. K.  van der Hucht \& B. Hidayat (Dordrecht: Kluwer),~229 (BP91)
\item   Breysacher, J., Moffat, A. F. J., \& Niemela, V. 1982, ApJ, 257, 116 
\item   Breysacher, J., \& Westerlund, B. E. 1978, A\&A, 67, 261
\item   Cant\'o J., Wilkin,  \& Raga, A. 1996, ApJ, 469, 729
\item   Charbonneau, P. 1995, ApJS 101 309
\item   De Marco, O. 2002, ASPC, 260, 517.
\item   Demers, H., Moffat, A.F.J, \& Marchenko, S.V. 2002,   ASPC 260, 563.
\item   Drissen, L., Crowther, P.A., Smith, L.J., Robert, C., Roy, J.-R.,  \& Hillier, D.J. 2001, ApJ 545, 484.
\item    Everett, M.E., \& Howell, S.B.  2001, PASP, 113, 1428
\item   Gayley, K. et al. 1997, ApJ 475, 786
\item   Georgiev, L. \& Koenigsberger, G. 2004a, A\&A, 423, 267.
\item   Georgiev, L. \& Koenigsberger, G. 2004b, IAU Symp. 215, {\it Stellar Rotation},   (ed) Maeder\& Eenens, p. 19
\item   Hoffmann, Stift \& Moffat 1978, PASP 90, 101.
\item   L\"uhrs, S. 1997, PASP 108, 504
\item   Kaufer, A., Schmid, H.M., Schweickhardt, J., Tubbesing, S. 2002,   in ``Interacting Winds from Massive Stars", ASP Conf. Ser. Vol. 260, 489.
\item   Koenigsberger, G., Auer, L.H., Georgiev,L. and Guinan, E. 1998a, ApJ, 496, 934
\item   Koenigsberger, G., Pe\~na, M.,  Schmutz, W. and Ayala, S. 1998b,  ApJ, 499, 889
\item   Koenigsberger, G., Georgiev, L., Barb\'a, R., Tzvetanov, Z., Walborn, N.R.,Niemela, V., Morrell, N., y Schulte-Ladbeck, R.  2000, ApJ, 542, 428
\item   Koenigsberger, G. 2004, RMAA, 40, 107
\item   Koenigsberger, G., Fullerton, A., Massa, D., Auer, L. 2006, AJ, 132, 1527
\item   Koenigsbeger, G. \& Moreno, E. 2007, {\it arXiv0705.1938}.
\item   De Marco, O.  2002, ASP Conf. Ser. 260, 517
\item   Marchenko, S.V., Moffat, A.F.J. \& Koenigsberger, G. 1994,  ApJ,  422, 810
\item   Marchenko, S.V., Moffat, A.F.J., Eenens, P.R.J., Cardona, O., Echevarria, J. Hervieux, Y. 1997,  ApJ, 485, 826 
\item   Massey, P., Parker, J.W., Garmany, C.D. 1989, AJ 98, 1305.
\item   Moffat, A.F.J., Marchenko, S.V., Bartzakos, P., Niemela, V.S., Cerruti, M.A.,Magalhaes, A.M.,  Balona, L., St.-Louis, N., Seggewiss, W., Lamontagne, R.  1998, ApJ 497, 896
\item   Naz\'e,  Y., Corcoran, M.F., Koenigsberger, G., Moffat, A.F.J. 2007, ApJLett, 658, 25.
\item   Nielsen,K.E., Corcoran, M.F., Gull, T.R., Hillier, D.J., Hamaguchi, K., Ivarsson, S., Lindler, D.J. 2007, ApJ 660, 669.
\item   Niemela, V.S. 1988, ASPCS 1, 381
\item   Niemela, V.S., Barb\'a, R.H., Morrell, N.I. and Corti, M. 1997, in ASP Conf Ser. 120, A. Nota and H.J.G.L.M. Lamers (eds), p.222.
\item   Pittard, J.M. 1998, MNRAS, 300, 479
\item   Pittard, J.M. 2007, ApJ, 660, L141.
\item   Schweickhardt, J. 2000, PhD Thesis, Ruprecht-Karls-Universit\"at, Heidelberg.
\item   Sterken C., Breysacher J. 1997, A\&A 328, 269
\item   Stevens, I.R., Blondin, J.M., Pollock, A.M.T. 1992, ApJ 386, 265.
\item   Stevens, I.R. 1993, ApJ, 404, 281.
\item   St.-Louis, N., Moffat, A.F.J., Marchenko, S.V., Pittard, J.M.  2005, ApJ, 628, 953.
\item   St.-Louis, N., Moffat, A.F.J., Marchenko, S.V., Pittard, J.M., Boisvert, P. 2006, ASPC,
 348, 121.
\item   Villar-Sbaffi, A., Moffat, A.F.J., St.-Louis, N. 2003, ApJ 590,483
\item   Walder, R. \& Folini, D.  2002, ASP Conf. Ser. 260, 595.

\end{itemize}

\begin{table*}[!t]\centering
\tablecols{4}
\setlength\tabnotewidth{0.60\textwidth}
\setlength{\tabcolsep}{0.9\tabcolsep}
\small
\caption{Summary of HD~5980 Estimated Parameters}
\begin{tabular}{llll}
\toprule
Parameter                    & 1979-81                & 1994       &1999-2005  \\
\midrule
$P_{orb}$ (days)             &   19.2654              &   $\cdots$              & $\cdots$ \\
Orbital inclination (deg)    &   88\tabnotemark{m}    &   $\cdots$              & $\cdots$ \\
a ($R_\odot$)                &  127\tabnotemark{b}    &   $\cdots$              & 143--157\tablenotemark{k} \\
e                            & 0.3                    &   $\cdots$              &0.30$\pm0.16$ \tablenotemark{k} \\
$M_{\rm A}$ (M$_\odot$)      &50\tabnotemark{b}       &   $\cdots$              & 58--79\tablenotemark{k} \\
$M_{\rm B}$ (M$_\odot$)      & 28\tabnotemark{b}      &   $\cdots$              & 51--67\tablenotemark{k}   \\
M$_V^{sys}$ (mag)            &  $-$7.3\tabnotemark{a} &$-$8.7\tabnotemark{e}    &$-$7.6\tabnotemark{h} \\
M$_V^A$ (mag)                &  $-$6.3                &$-$8.6\tabnotemark{e}    & $-$6.5  \\
M$_V^B$ (mag)                &  $-$5.8                &  $\cdots$               &$\cdots$   \\
M$_V^C$ (mag)                &  $-$6.1                &  $\cdots$               &$\cdots$  \\
$R_{{\rm A}}$ (R$_\odot$)    &  21\tabnotemark{a}     &  $> 160$\tabnotemark{d} &  $\cdots$ \\
                             &   $\cdots$             &  280\tabnotemark{g}     & $\cdots$ \\
$R_{{\rm  B}}$ (R$_\odot$)   &  15\tabnotemark{a}     &   $\cdots$              &$\cdots$ \\
$R_{{\rm env B}}$ (R$_\odot$)&  30--40\tabnotemark{a} &  $\cdots$               &$\cdots$ \\
$v_\infty^A$  (km/s)         &  $\cdots$              & 300--1700               & 1740--2200: \\
                             & $\cdots$                & 500\tabnotemark{g}      &  $\cdots$   \\
$v_\infty^B$  (km/s)         & 3100:                   & $\cdots$                &  2600--3100:  \\
$v_\infty^C$  (km/s)         &    $\cdots$             &  $\cdots$               &1700:: \\
N[He]/N[H]$^{aa}$$_{\rm Non-LTE}$  & $\cdots$          &  0.4 \tabnotemark{e}     &$\cdots$  \\
$\dot{M}_A$ (M$_\odot$/yr)   & $\cdots$                            & $10^{-3}  $\tabnotemark{e}&$\cdots$  \\
$\dot{M}_B$ (M$_\odot$/yr)   & 2$\times$10$^{-5}  $\tabnotemark{m}   &                         &         \\
$L_A$  (L$_\odot$)           &    $\cdots$             &  $3\times 10^6  $\tabnotemark{e} & $\cdots$ \\
                             &    $\cdots$       &  $10^7  $\tabnotemark{g}  &$\cdots$  \\
$T_*^A$ $^\circ$K            &53,000\tabnotemark{c} & 21,000\tabnotemark{d} &$\cdots$  \\
                             & $\cdots$          & 35,500\tabnotemark{e}   & $\cdots$  \\
Spectrum ({\rm Star A})      & WN3\tabnotemark{b}& $B1.5Ia^+$\tabnotemark{d}  &  WN6\tabnotemark{d}\\
Spectrum ({\rm Star B})      &WN4\tabnotemark{b} &  WN4:                    &  WN4:    \\
Spectrum ({\rm Star C})      &$\cdots$           & $\cdots$                 & O4-6\tabnotemark{f}\\
                             & $\cdots$          &  $\cdots$                & O7I \tabnotemark{n}\\
$V_{rot}^A \, sin~i$ (km/s)  &$\cdots$           &  $\cdots$                 & 250::\tabnotemark{j}  \\
$V_{rot}^C \, sin~i$ (km/s)  &$\cdots$           &  $\cdots$                 & 75\tabnotemark{i}  \\
P$_{orb}$(C) (days)         &  $\cdots$         &  $\cdots$                 & 96.5 \tabnotemark{s}   \\
\bottomrule
\tabnotetext{aa}{Abundance ratio by number (0.63 by mass); \quad
\textsuperscript{a}Breysacher \& Perrier 1991; \quad \textsuperscript{b}Niemela et~al. 1997;\quad
\textsuperscript{c}Assuming $L_{\rm star A}$ ($L_\odot$)=const and
$R_{\rm star A}$/$R_\odot$=21; \quad
\textsuperscript{d}Koenigsberger  et~al.\@    1998a; \quad
\textsuperscript{e}December 30,
Koenigsberger  et~al.\@    1998b; \quad
\textsuperscript{f}Koenigsberger  et~al.\@    2000; \quad
\textsuperscript{g}Drissen  et~al.\@    2001; \quad
\textsuperscript{h}S. Duffau, assuming m-M=19.1 mag;\quad
\textsuperscript{i}Koenigsberger et~al.\@    2001;\quad
\textsuperscript{j}Georgiev \& Koenigsberger 2004b;\quad
\textsuperscript{k} This paper;\quad
\textsuperscript{m} Moffat et al. 1998; \quad
\textsuperscript{n} Koenigsberger, Fullerton, Massa \& Auer 2006;\quad
\textsuperscript{s} Schweickhardt, J. 2002.
}            
\end{tabular}
\vspace*{0.7\baselineskip}
\end{table*}

\begin{table*}[!t]\centering
\tablecols{12}
\setlength\tabnotewidth{0.60\textwidth}
\setlength{\tabcolsep}{0.9\tabcolsep}
\small
\caption{FEROS 1999 data}
\begin{tabular}{lrrrrrrrrrrr}
\toprule
spectrum&  MJD & phase & NIV & 4057.76  &  & HI &4101.71& & HeII & 5411.52&    \\
        &      &       & RV  & FW    &$-$EW & RV  & FW    &$-$EW& RV  & FW    &$-$EW\\
        &      &       &km/s &km/s   & \AA  &km/s &km/s   & \AA &km/s &km/s   & \AA \\
\midrule
 78781 & 51374.418 & 0.475 & 13 & 685 &3.83 & 76 & 1105 &7.94 & 203 & 907 & 14.79  \\    
 79411 & 51375.422 & 0.527 &-10 & 694 &4.46 & 63 & 1221 &8.99 & 221 & 993 & 16.04  \\                          
 80871 & 51379.402 & 0.734 &-50 & 689 &2.99 & 56 & 1515 &9.03 & 201 &1217 & 15.62  \\        
 81451 & 51380.398 & 0.786 &-55 & 685 &3.05 & 53 & 1546 &8.80 & 211 &1306 & 15.21  \\                
 82121 & 51381.402 & 0.838 &-23 & 705 &2.79 & 47 & 1538 &8.58 & 218 &1249 & 14.84  \\                      
 83181 & 51382.352 & 0.887 &-0.5 &711 &3.03 & 66 & 1505 &8.91 & 227 &1213 & 15.18  \\    
 83811 & 51383.355 & 0.939 & 50 & 679 &3.14 & 97 & 1527 &9.08 & 251 &1188 & 14.11  \\                    
 84511 & 51384.309 & 0.989 &116 & 697 &3.39 &113 & 1316 &8.51 & 226 &1066 & 13.48  \\             
 85181 & 51385.332 & 0.042 &197 & 663 &3.72 &187 & 1072 &9.41:& 317 & 769 & 13.27  \\                  
 85771 & 51386.391 & 0.097 &229 & 760 &3.46 &210 & 1297 &8.55 & 340 &1023 & 14.22  \\                 
 86321 & 51388.340 & 0.198 &187 & 810 &3.98 &233 & 1553 &8.54 & 404 &1185 & 12.99  \\                
 86701 & 51389.383 & 0.252 &138 & 752 &3.53 &230 & 1424 &8.02 & 399 &1049 & 13.06  \\               
 87211 & 51390.328 & 0.301 &118 & 647 &3.46 &143 & 1373 &8.41 & 328 &1070 & 14.01  \\              
 87751 & 51391.332 & 0.353 & 92 & 624 &2.68 & 66 & 1215 &7.42 & 207 & 946 & 13.46  \\     
 88931 & 51392.328 & 0.405 & 49 & 609 &3.28 & 43 & 1170 &8.48 & 175 & 872 & 15.55  \\            
 89511 & 51393.371 & 0.459 & 23 & 658 &3.57 & 55 & 1118 &8.22 & 195 & 880 & 14.91  \\         
 90131 & 51394.391 & 0.512 &0.5 & 671 &3.76 & 54 & 1136 &8.43 & 210 & 921 & 15.82  \\          
   \bottomrule
  \end{tabular}
\end{table*}

\begin{table*}[!t]\centering
\tablecols{12}
\setlength\tabnotewidth{0.60\textwidth}
\setlength{\tabcolsep}{0.9\tabcolsep}
\small
\caption{FEROS 2005 data}
\begin{tabular}{lrrrrrrrrrrr}
\toprule
spectrum&  MJD & phase & NIV & 4057.76  &  & HI &4101.71& & HeII & 5411.52&    \\
        &      &       & RV  & FW    &$-$EW & RV  & FW    &$-$EW& RV  & FW    &$-$EW\\
        &      &       &km/s &km/s   & \AA  &km/s &km/s   & \AA &km/s &km/s   & \AA \\
\midrule
0617801 & 53538.385& 0.799 &-29 & 760 &2.56 & 21 & 1860 &5.82 & 211 &1622 & 15.33  \\
0617531 & 53538.404& 0.800 &-31 & 756 &2.63 & 12 & 1903 &5.80 & 221 &1643 & 14.56  \\
0620221 & 53541.384& 0.955 & 78 & 800 &2.70 &136 & 1692 &5.31 & 278 &1454 & 12.96  \\
0620344 & 53541.403& 0.955 & 73 & 810 &2.99 &120 & 1608 &5.62 & 291 &1497 & 13.38  \\
0710740 & 53561.367& 0.992 & 98 & 745 &3.13 &141 & 1409 &6.37 & 281 &1099 & 14.07  \\
0710950 & 53561.386& 0.993 & 98 & 729 &3.11 &143 & 1395 &5.98 & 284 &1068 & 14.83 \\
0725580 & 53576.288& 0.767 &-10 & 817 &2.65 & 91 & 2003 &5.13 & 270 &1705 & 15.39  \\
0725630 & 53576.307& 0.767 &-3  & 828 &2.76 & 77 & 1921 &5.82 & 266 &1654 & 15.03  \\
0921331 & 53635.000& 0.814 &-19 & 936 &2.32:& 23 & 1952 &4.62 & 216 &1747 & 15.36  \\
0922990 & 53635.019& 0.815 &-2  & 955 &2.36 & 56 & 2026 &5.57 & 202 &1757 & 14.99  \\
0925220 & 53638.253& 0.983 &118 & 694 &2.75 &174 & 1452 &5.11 & 295 &1239 & 13.38  \\
0925481 & 53638.272& 0.984 &113 & 758 &2.54 &178 & 1427 &4.97 & 285 &1192 & 13.24 \\
0928220 & 53641.153& 0.133 &249 & 986 &2.46 & ?? & ??   &4.60 & 420 &1629 & 13.50 \\
0928161 & 53641.173& 0.134 & 260& 921 &2.40 & ?? &  ??  & 4.82& 420 &1627 & 13.46 \\
1022700 & 53665.026& 0.372 & 85 & 657 &2.53 &16- & 1290 & 5.72 &159 &1088 & 15.75 \\
1022731 & 53665.045& 0.374 & 86 & 666 &2.53 &0.1 & 1288 & 5.59 &153 &1048 & 15.92 \\
    \bottomrule
  \end{tabular}
\end{table*}

\begin{table*}[!t]\centering
\tablecols{3}
\setlength\tabnotewidth{0.60\textwidth}
\setlength{\tabcolsep}{0.9\tabcolsep}
\small
\caption{EMMI data}
\begin{tabular}{lrr}
\toprule
spectrum &   MJD & phase \\
\midrule
0619 &  53541.389 & 0.955   \\
0630 &  53552.429 & 0.528   \\
0710 &  53561.400 & 0.994   \\
0921 &  53634.999 & 0.814   \\
0924 &  53638.238 & 0.982   \\
0927 &  53641.176 & 0.135   \\
1010 &  53654.110 & 0.806   \\
1012 &  53656.164 & 0.913   \\
    \bottomrule
  \end{tabular}
\end{table*}

\begin{table*}[!t]\centering
\tablecols{9}
\setlength\tabnotewidth{0.60\textwidth}
\small
\caption{Photospheric absorptions 2005}
\begin{tabular}{lrlrlrlrl}
\toprule
          &$\phi=$&0.8&$\phi=$&0.0&$\phi=$&0.13&$\phi=$&0.37\\
Atomic ID & RV & FW & RV & FW& RV & FW& RV & FW  \\
\midrule
He I 3819.60 & 142&175 &190&286& --&--   &??&??      \\
H 9 3835.38 & 167&260 &170&214& 186&224 &140&267      \\
          &  --&--  &-- &-- &-120&468 &   &        \\
He I 3888.64\tabnotemark{c} & 208&282 &198&207\tabnotemark{a}& 163&264&195&201  \\
He I 4026.18 & 182& 178 &153&119 &154&171&144&127    \\
          & --  --&  &--&--  &-324&554&-- &--    \\
H I 4101.76: & 194& 265 &176&213& 157&239& 156&221    \\
          & -- &--  &-- &---&-314\tabnotemark{b}&579& --& --    \\
He II 4199.84\tabnotemark{d}&174& 278 &176&164 &171&176 &166&186    \\
           &-- & --  &-- &-- &-249&547 &-- &--    \\
H I 4340.49  & 163& 220 &135&232& 141&247 &141&185    \\
          & -- & --  &   &   &-351&406 & --  &  --     \\
He I 4387.93 & 149& 276 &175&137& 188&326 &170&189    \\
He I 4471.50:& 168& 185 &167&152& 161&158 &155&132    \\
He II 4541.59&190& 181 &192&126& 165&137 &168&183    \\
             &   &     &   &   & -249 &545    &   &        \\
H I 4861.30   &176& 287 &113:&311&141&257 &188&326    \\
He I 4921.93  &165& 207 &173&163 &156&166 &155&178    \\
He I 5015.68  &181& 144 &179&148 &173&160 &167&123\tabnotemark{b}   \\ 
He II 5411.52& 181& 178 &197&128& 177&141 &189&125       \\
           & -- & -- &-142&145&-323&400 & --   &  --       \\
O III 5592.4 & 166& 150& 156&149& 168&134 &148&117    \\
C IV 5801.3  & 178& 177& 203&180& 165&141 &182&170   \\
C IV 5812.0  & 177& 132& 200&117& 199&136 & --  &--      \\
He I 5875.8   & 173& 220& 164&190& 162&180 &162&160    \\
He I 6678.15  & 436& 181& 187&113& 122&250 &-- &--      \\
           & --& --  &-145&324& 419&151 &-- &--    \\
He I 7065.71  & 150& 250& 154&199& 147&234 &141&129    \\
           &  --& -- & -- &--&- 549&569 & --& --    \\
N IV 7111.28 & 152 & 93& 170&105& 129& 73 & 136& 74    \\
             &   8& 156& -- &-- & -- &--   &-- & --    \\
N IV 7122.98 & 277&  110& --&-- &-- & --& --&--    \\
             &  16&   81&   &-- &--& --& --&--    \\
$<>$         & 172&  --  & 173 & --& 161 & -- & 161 & -- \\
s.d.         &$\pm$16&  --  &$\pm$22&  --  &$\pm$19&  --  &$\pm$18&  --  \\
\bottomrule
\tabnotetext{}{  \quad
 \textsuperscript{a} filled-in by emission; \quad \textsuperscript{b} very
clear \quad \textsuperscript{c} blended with H I 3889.05  \quad
\textsuperscript{c} blended with N III 4200.  \quad
}
  \end{tabular}
\vspace*{0.7\baselineskip}
\end{table*}

\begin{table*}[!t]\centering
\tablecols{6}
\setlength\tabnotewidth{0.60\textwidth}
\setlength{\tabcolsep}{0.9\tabcolsep}
\small
\caption{Average differential photometry: Danish telescope}
\begin{tabular}{lrrrrr}
\toprule
  $<$MJD$>$ & $<$phase$>$ & Num. & $<$m$_v$-C$>$ &  $<\sigma_{instr}>$ & s.d.   \\
\midrule
  53561.422  &0.995 & 20  & -1.244 & 0.008  &    0.014\\
  53562.352  &0.043 & 18  & -1.397 & 0.007  &    0.008\\
  53564.352  &0.147 & 12  & -1.501 & 0.008  &    0.006\\
  53568.305  &0.352 &  7  & -1.234 & 0.006  &    0.003\\
  53576.367  &0.771 & 43  & -1.423 & 0.008  &    0.006\\
  53684.203  &0.368 &172  & -1.186 & 0.004  &    0.013\\
  53685.105  &0.415 & 30  & -1.375 & 0.003  &    0.004\\
  53687.207  &0.524 & 41  & -1.440 & 0.005  &    0.007\\
  53689.191  &0.627 & 95  & -1.442 & 0.006  &    0.006\\
  53693.184  &0.834 & 20  & -1.452 & 0.006  &    0.007\\
   \bottomrule
 \end{tabular}
\end{table*}

------------------------


\onecolumn
        
\begin{figure}[!t]
\includegraphics[width=\columnwidth]{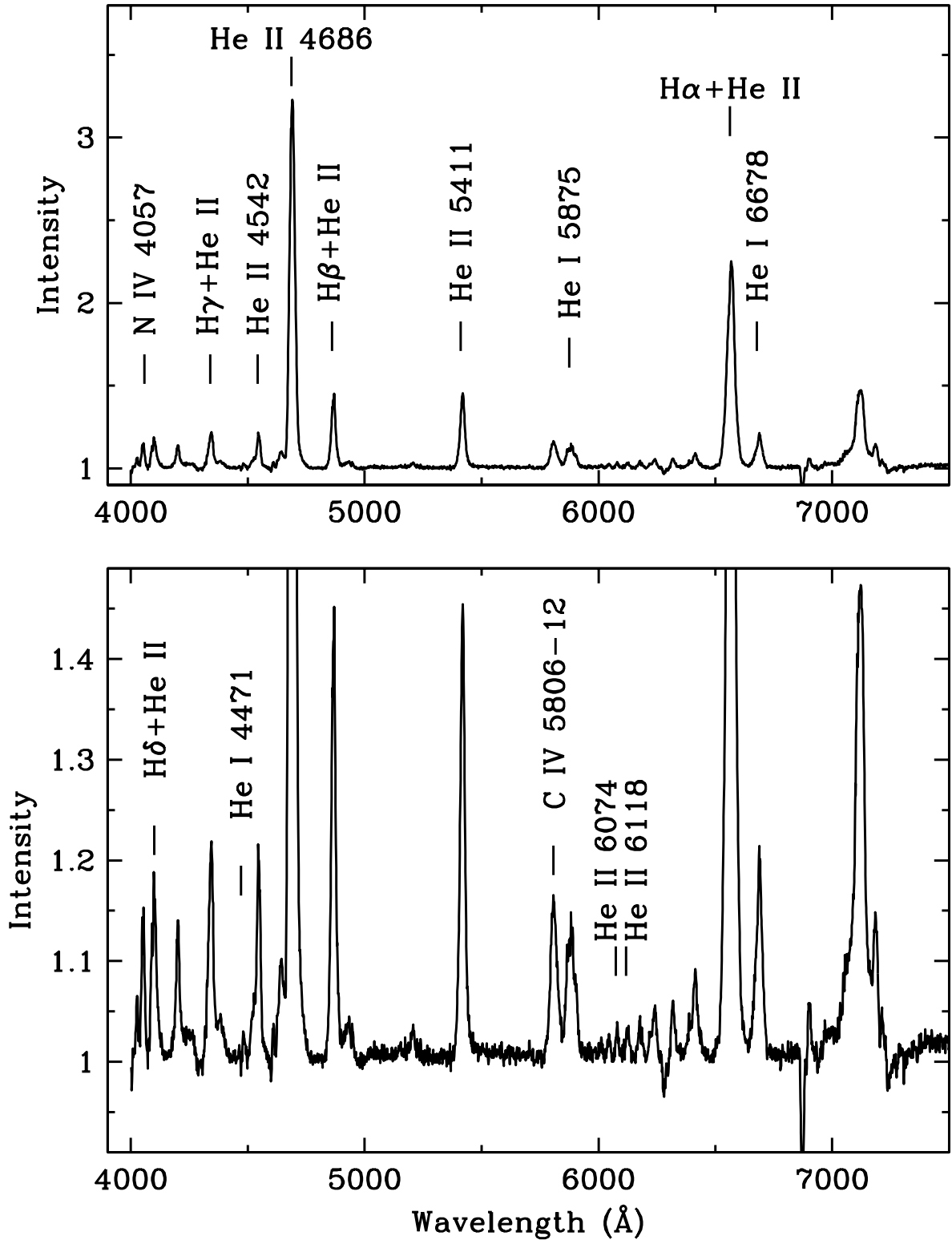}  
\caption{Average $\phi\sim$0 low dispersion spectrum obtained with EMMI in 2005
and rectified.
The strongest line in the spectrum is He II 4686 \AA.  Other lines that are
mentioned in the text are labeled with their primary identifications. The same 
spectrum is amplified in the lower panel, to display  more clearly the weaker lines. 
}
\end{figure}

\begin{figure}[!t]
\includegraphics[width=\columnwidth]{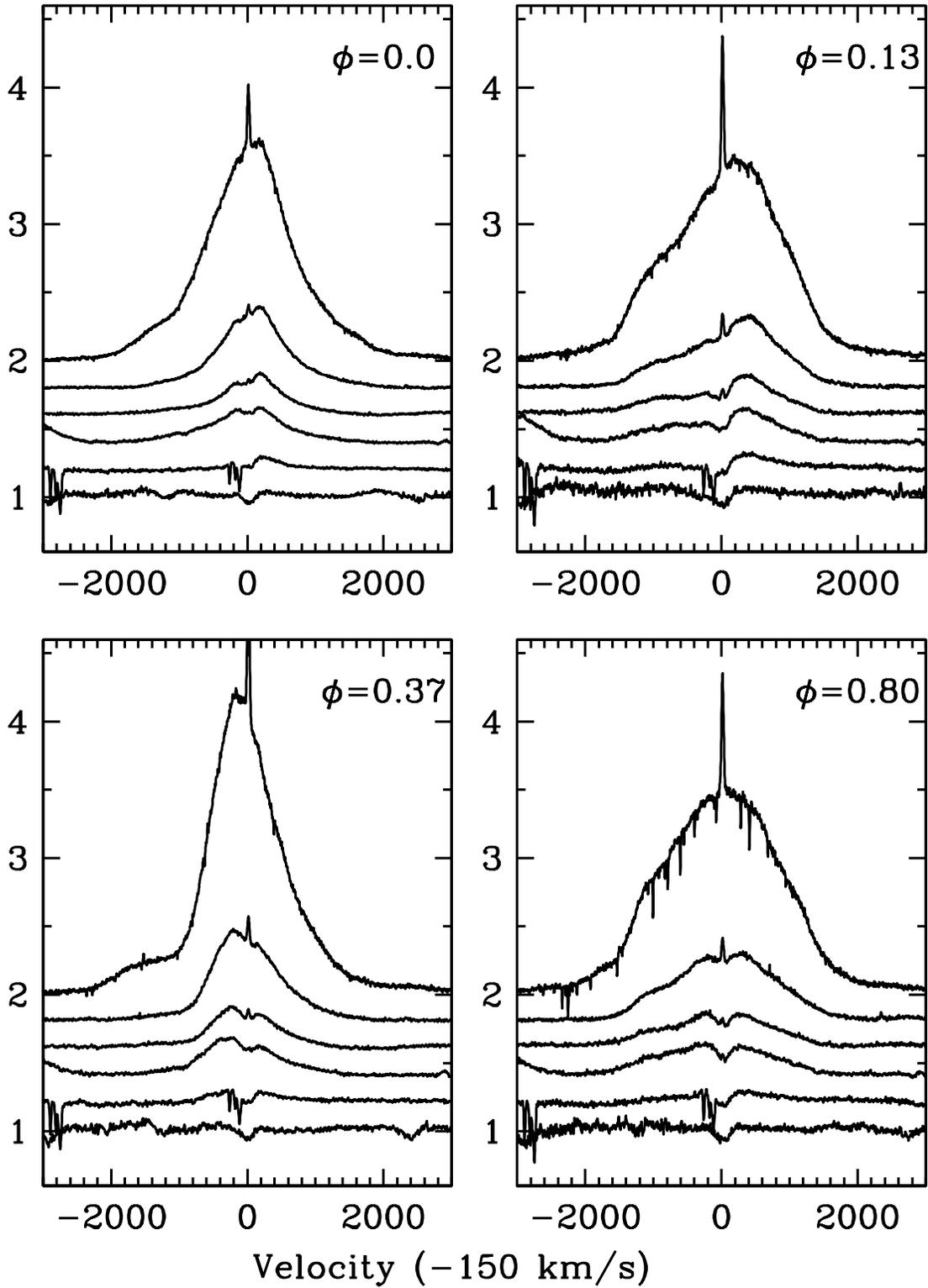}   
\caption{H I (+He II) line profiles at H$\alpha$, H$\beta$, H$\gamma$, H$\delta$, H$\epsilon$
and H9 3835 \AA\ from FEROS 2005 data. The spectra are averages for each of the orbital phases
$\phi=$0.0, 0.13, 0.37 and 0.80, and  the spectra at $\phi=$0.0 and 0.37 are renormalized to 
account for the corresponding eclipses. The velocity scale is corrected for the  +150 km/s systemic 
velocity of the SMC. Sharp emission spikes, strongest at H$\alpha$, are nebular emission.
}
\end{figure}

\begin{figure}[!t]
\includegraphics[width=\columnwidth]{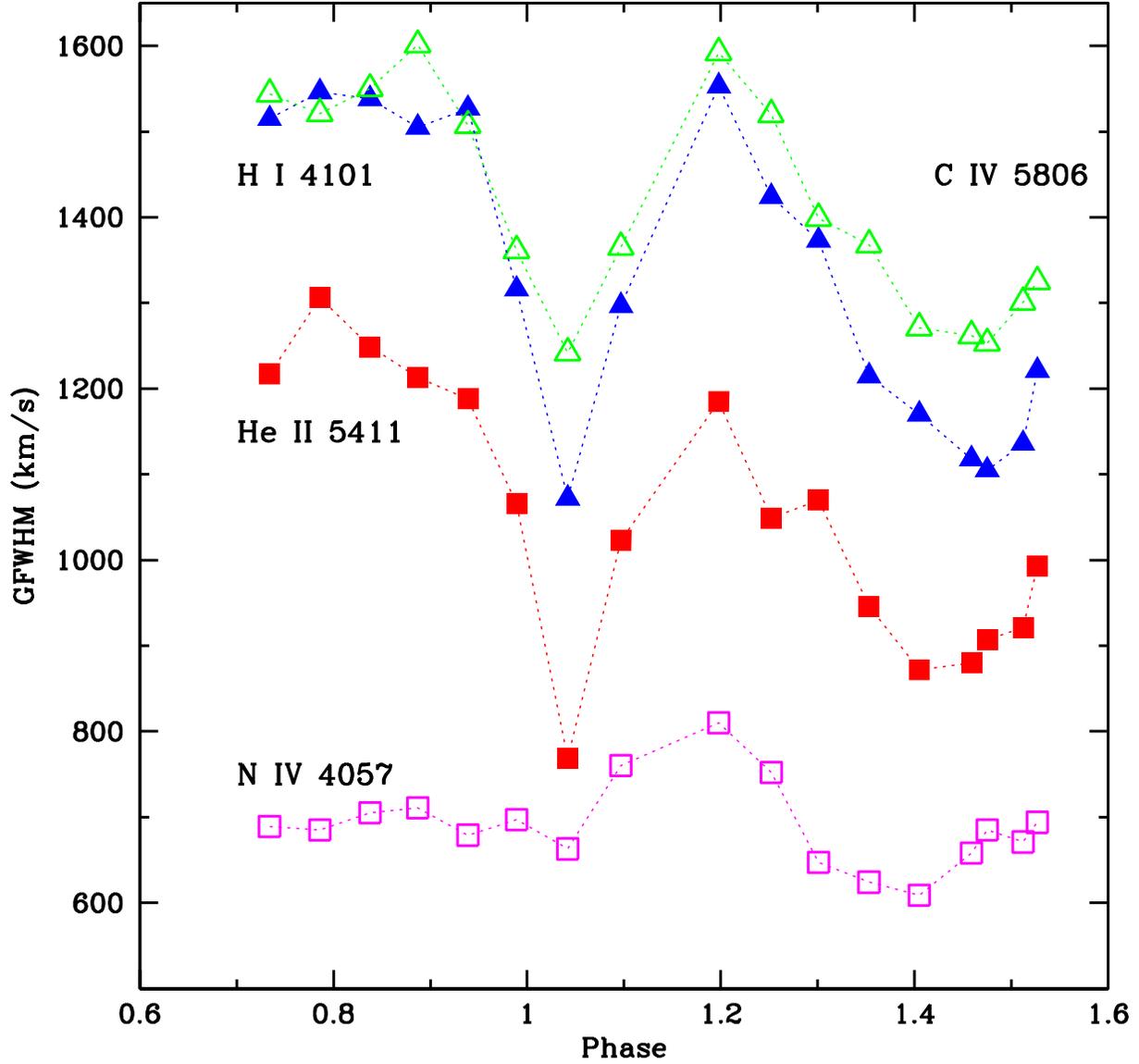}      
\caption{Full width at half maximum intensity measured with Gaussian fits to the
line profiles of NIV 4057 (open squares), He II 5411 (filled-in squares), C IV 5806
(open triangles), H I 4201 (filled-in triangles) in the 1999 spectra.  All lines except N IV display
the sudden decrease in width when {\it star A} is ``in front" ($\phi\sim$0). A more
gradual decrease is observed at the opposite eclipse ($\phi\sim$0.36).  This is the same behavior
seen in the late 1970's.  The dotted line connects the data points.   Data at
eclipse phases correspond to the renormalized spectra.
}
\end{figure}

\begin{figure}[!t]
\includegraphics[width=\columnwidth]{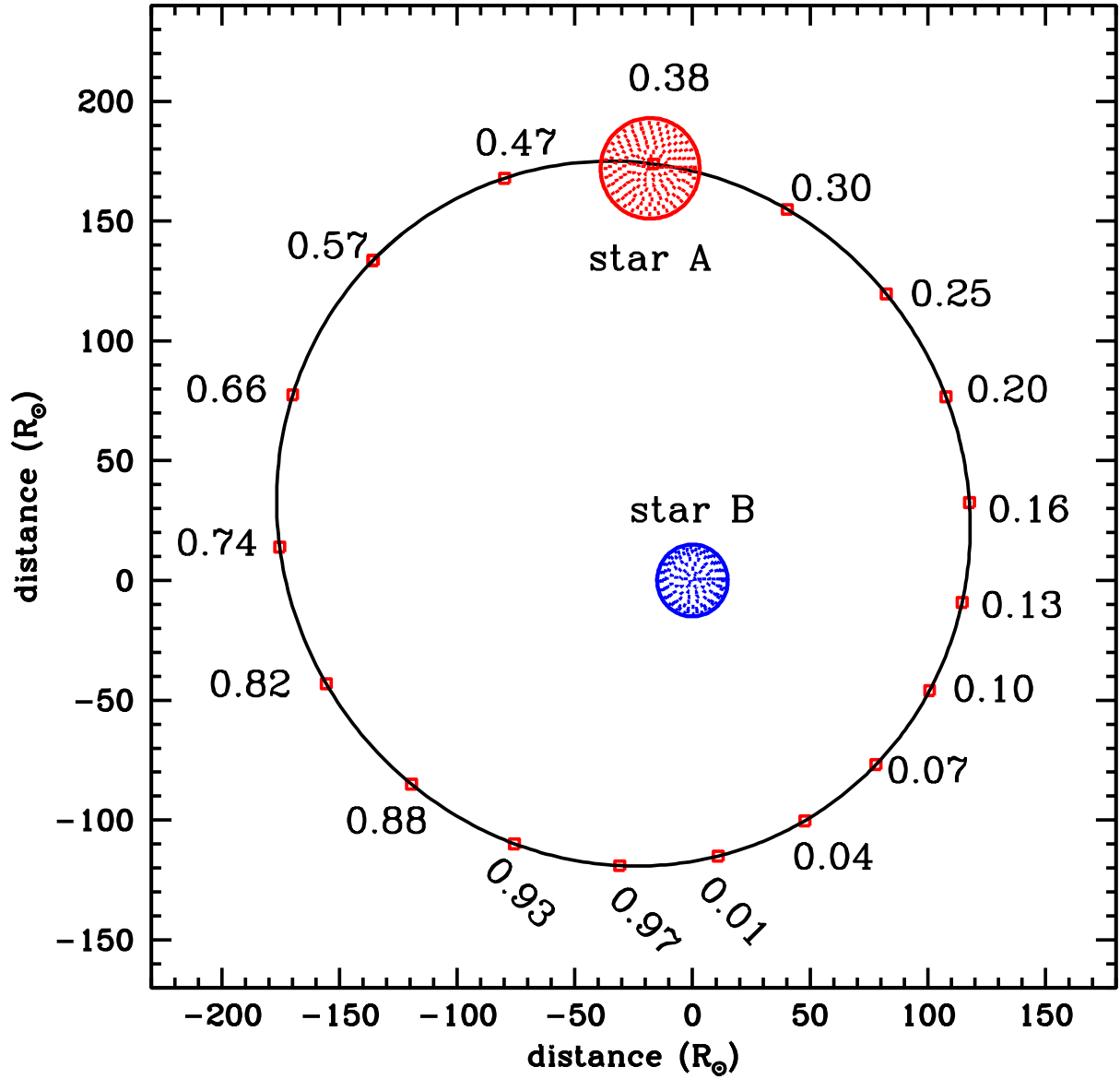}       
\caption{Schematic representation of the orbit indicating orbital phases in
a frame of references in which {\it star B} is at the origin.  For this representation,
the masses M$_A=$70 M$_\odot$ and M$_B=$54 M$_\odot$ (see Section 4), and
the eccentricity and argument of periastron are, respectively, e=0.3 and 
$\omega=$133$^\circ$; and R$_A=$21 R$_\odot$, R$_A=$15 R$_\odot$ from BP91.  The observer 
is located off the page at the bottom of the figure.
}
\end{figure}

\begin{figure}[!t]
\includegraphics[width=\columnwidth]{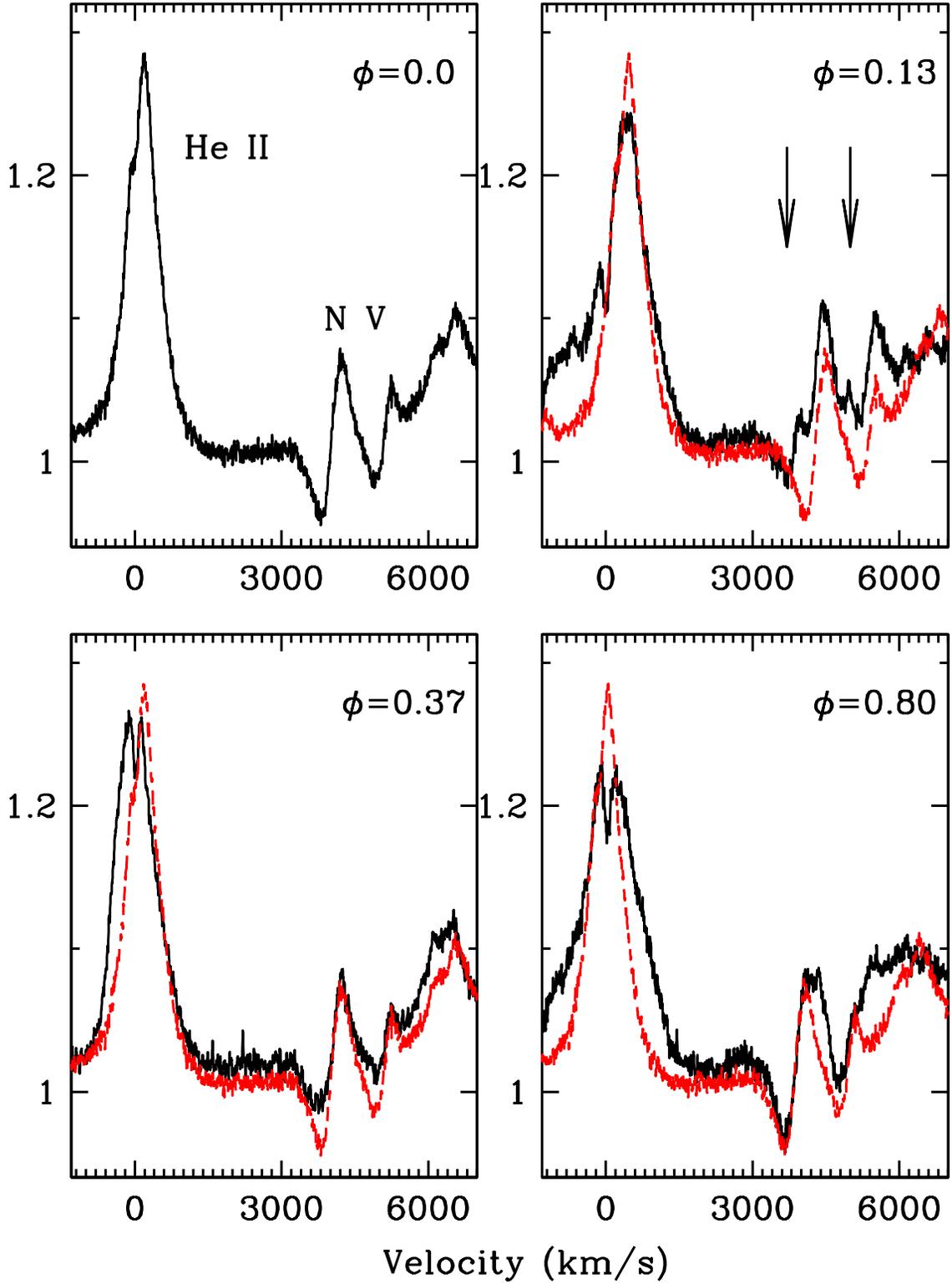}             
\caption{Line profiles at $\phi=$0.0 (top left) used as a template
to compare with profiles at $\phi=$0.13 (top right), 0.37 (bottom left), 
and 0.80 (bottom right).  The template is shifted by +275 km/s (top panel, right)
and -150 km/s (bottom panel, right). Spectra at $\phi=$0.0 and 0.37 are renormalized 
to account for the eclipses. The arrows point to excess emission that appears to be 
associated with {\it star B}.  The velocity scale is centered on  the 
He II $\lambda_0=$4541.59 line, corrected for SMC motion ($+$150 km/s).
}
\end{figure}

\begin{figure}[!t]
\includegraphics[width=\columnwidth]{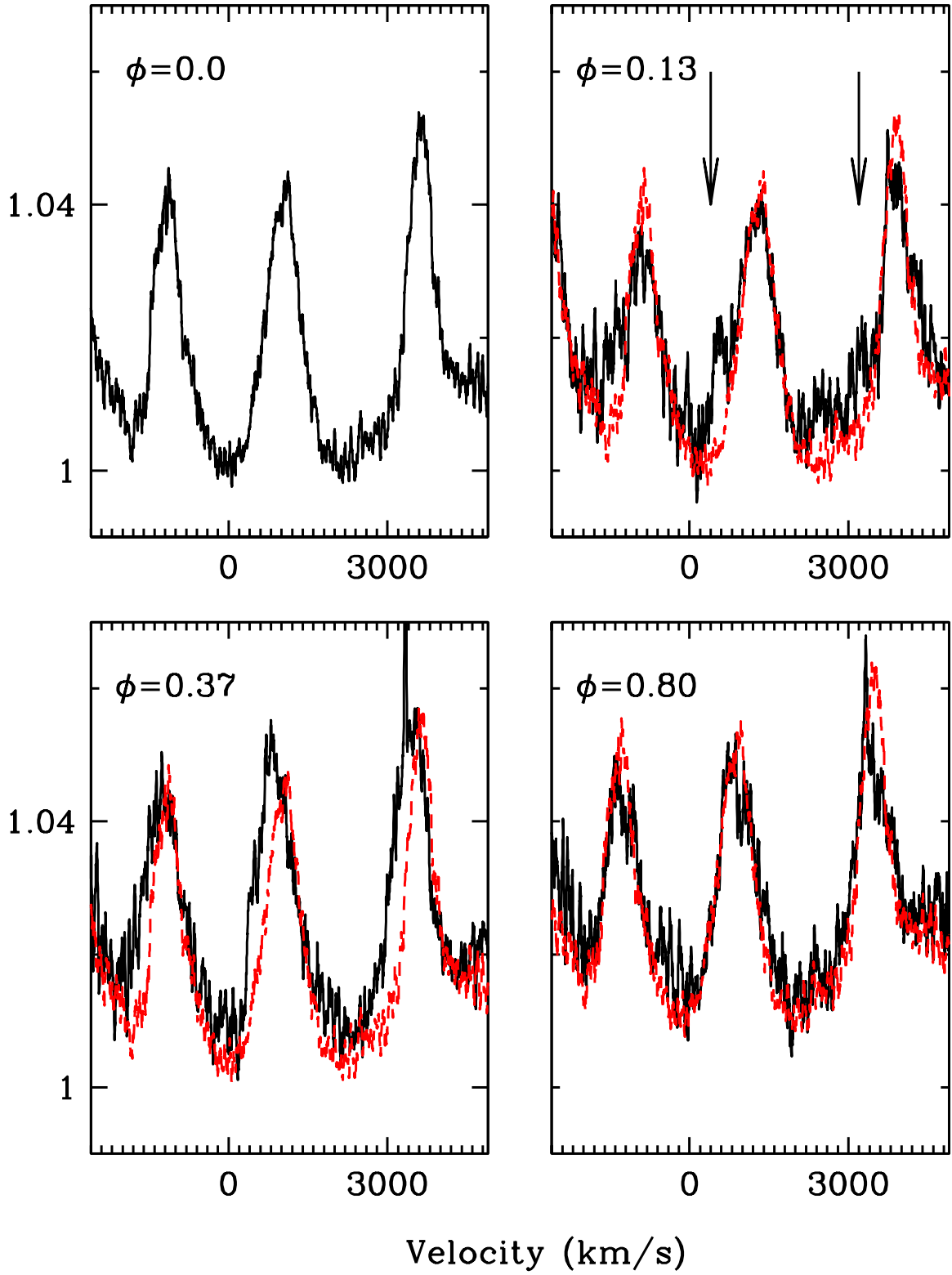}          
\caption{Same as previous figure, but for 3 of the He II lines
in the 6000-6200 \AA\ region.  V=0 km/s corresponds to $\lambda_0=$6100. \AA.
}
\end{figure}

\begin{figure}[!t]
\includegraphics[width=\columnwidth]{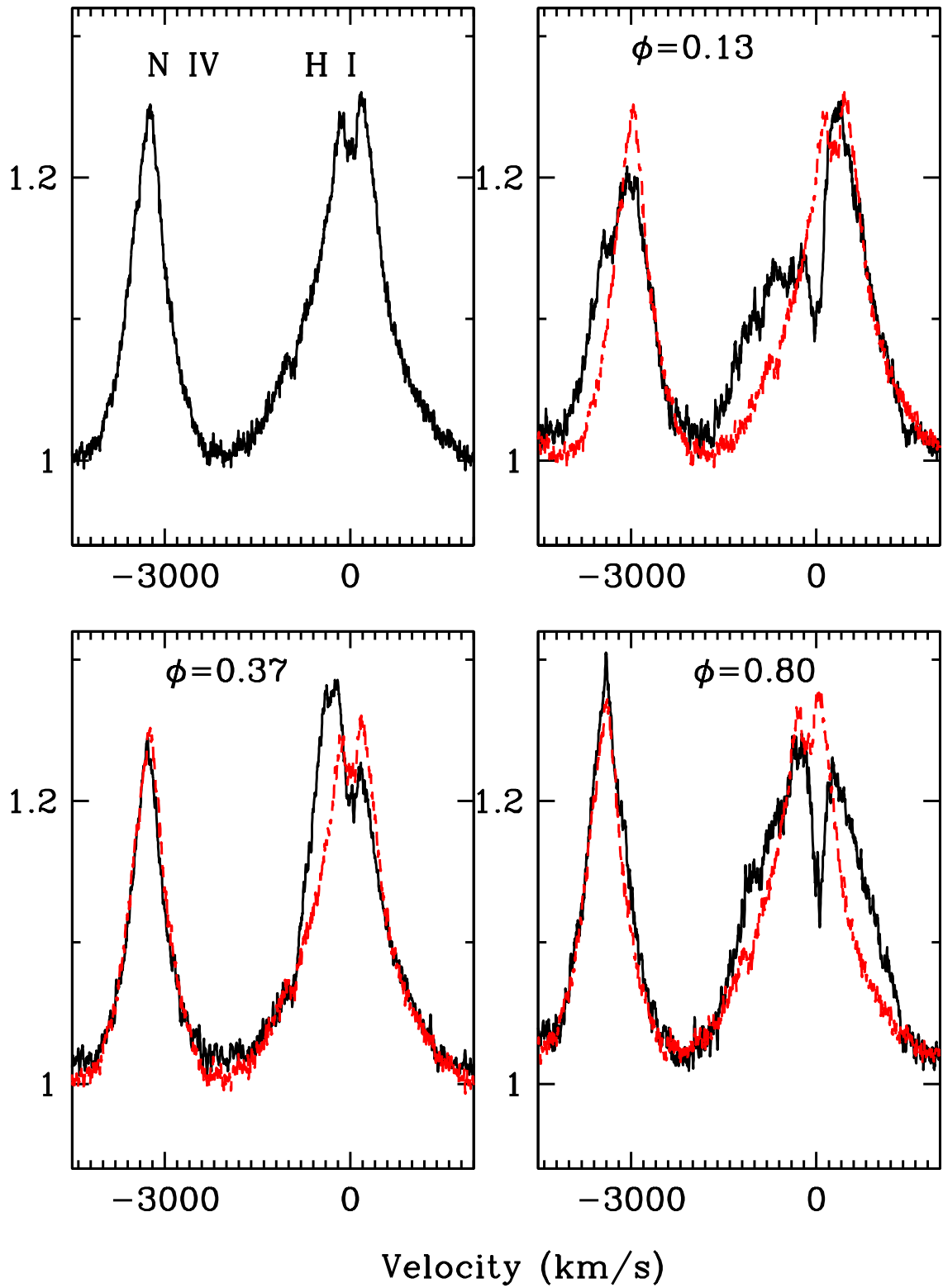}       
\caption{Same as previous figure, but for N IV 4057  and H I 4100 \AA.
Velocity scale is centered on $\lambda$ 4101.73 \AA.
}
\end{figure}

\begin{figure}[!t]
\includegraphics[width=\columnwidth]{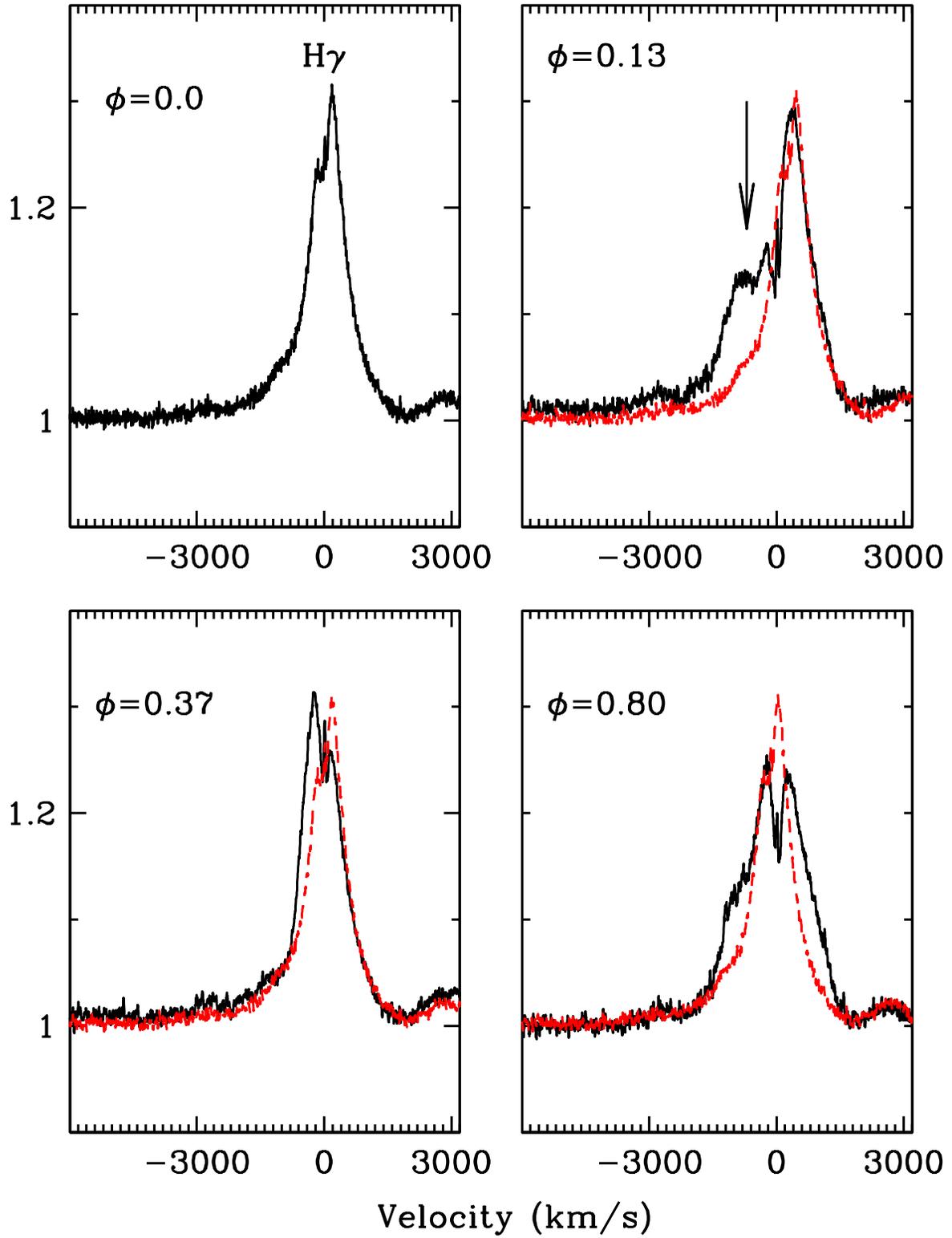}           
\caption{Same as previous figure, but for H$\gamma$.Velocity scale is centered 
on $\lambda$ 4340.45 \AA.
}
\end{figure}

\begin{figure}[!t]
\includegraphics[width=\columnwidth]{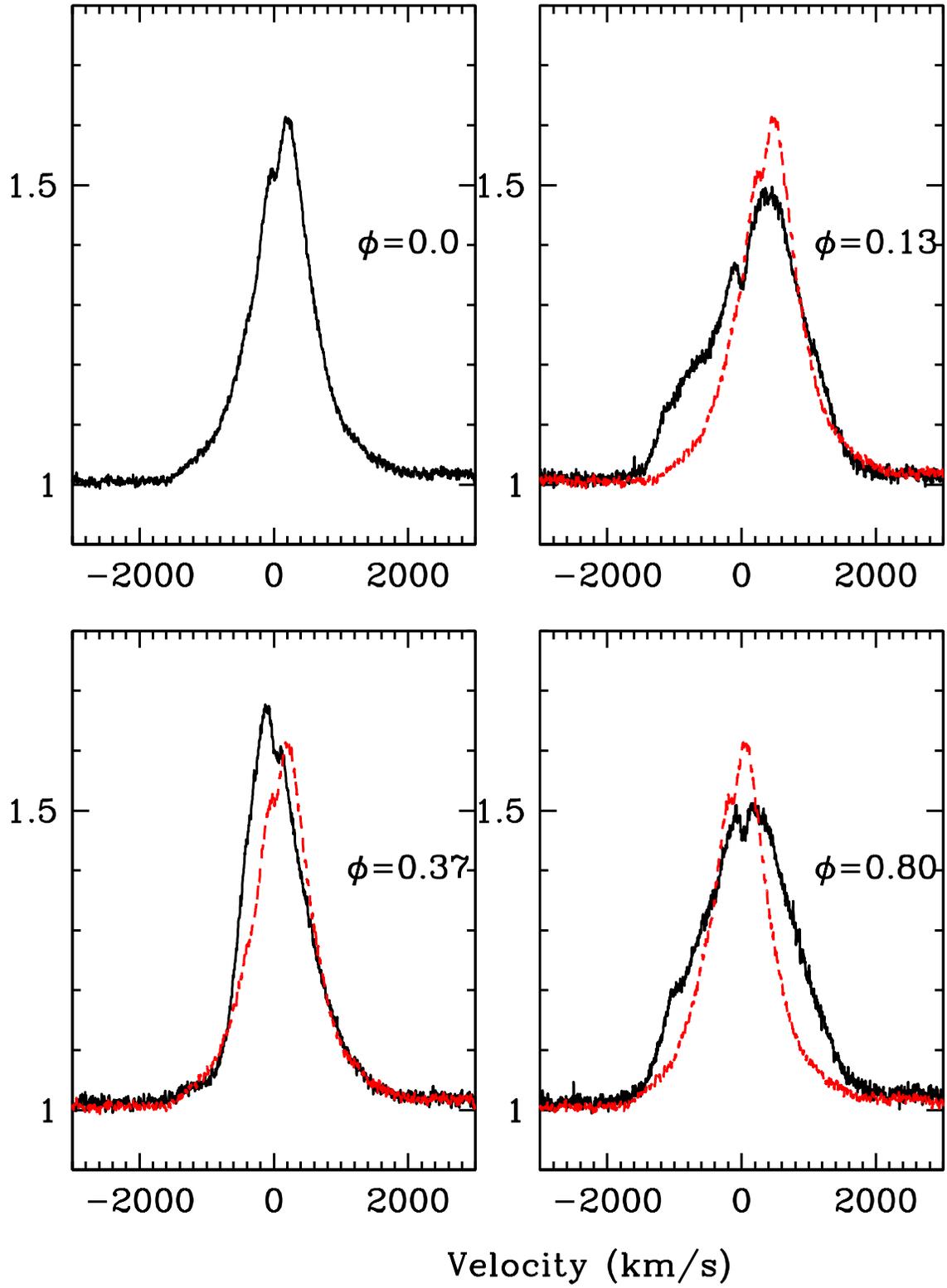}             
\caption{Same as previous figure, but for He II 5411 \AA. 
Velocity scale is centered on $\lambda$ 5411.52 \AA.
}
\end{figure}

\begin{figure}[!t]
\includegraphics[width=\columnwidth]{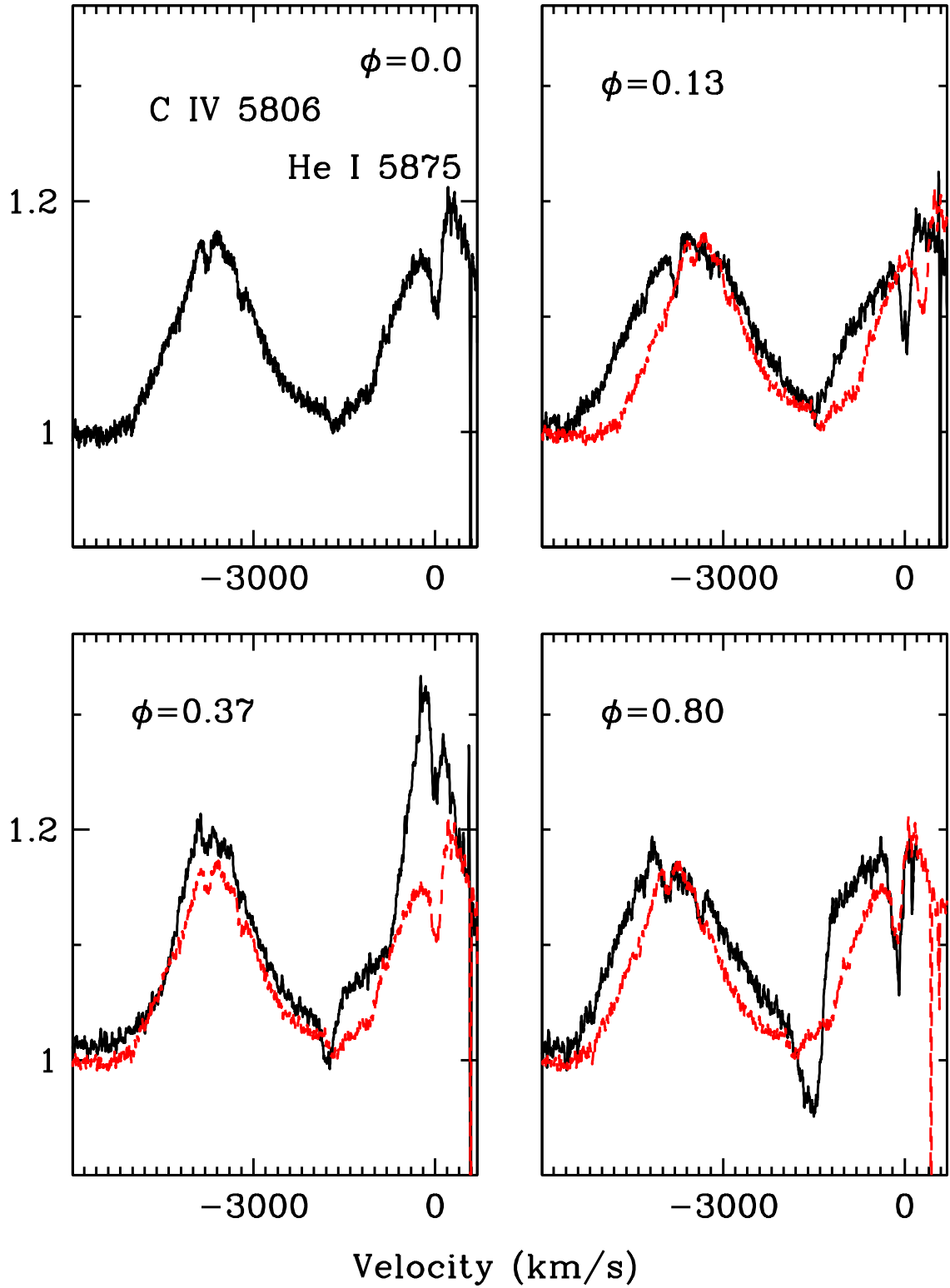}               
\caption{Same as previous figure, but for He I 5875.80 \AA.
}
\end{figure}

\begin{figure}[!t]
\includegraphics[width=\columnwidth]{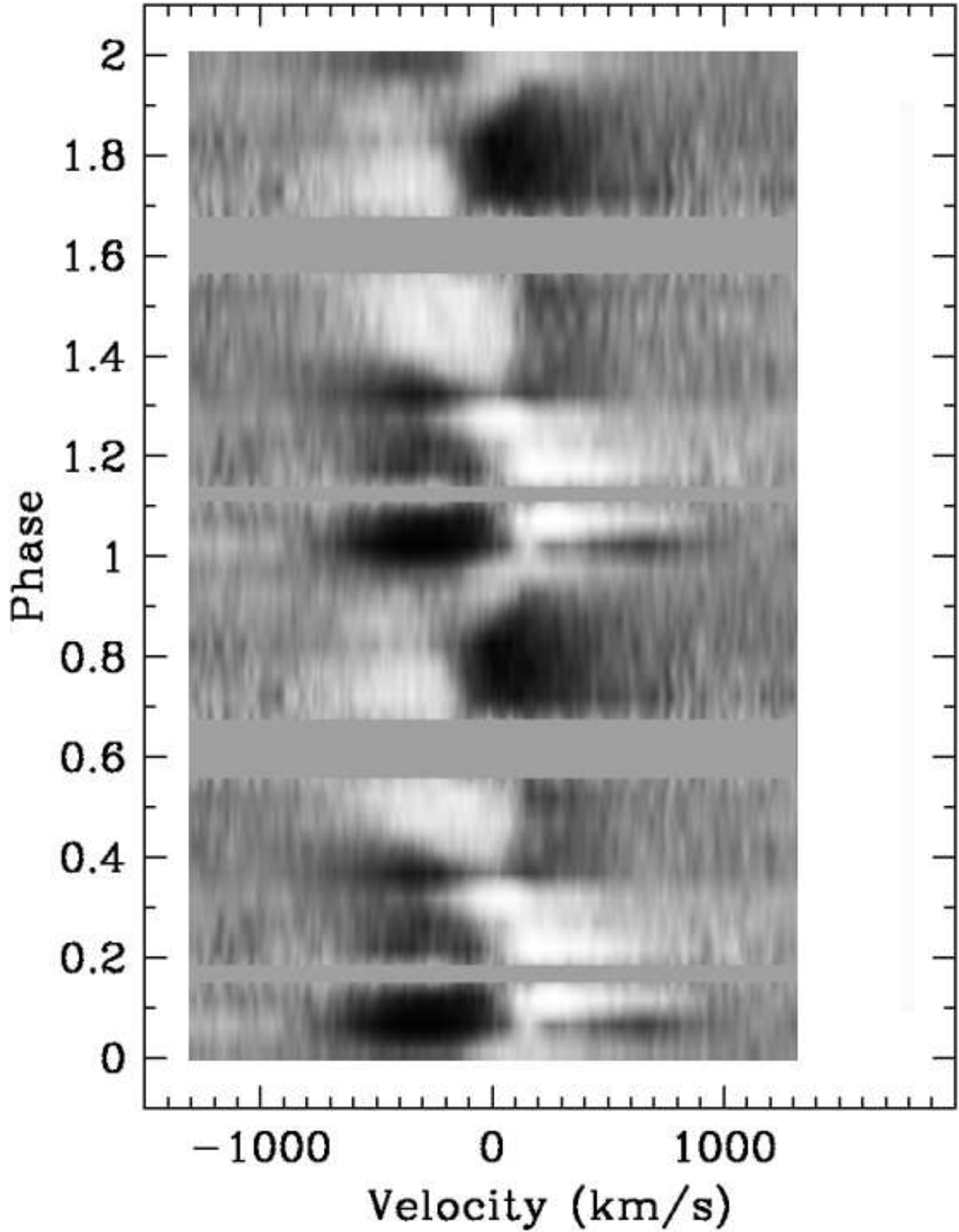}          
\caption{Grey scale representation of residuals  for N IV 4057. Phase increases from           
the bottom to the top, the same data being plotted twice. White corresponds to positive residuals, 
black to negative. Residuals are the difference between the nearly equally-spaced (in phase) 1999 
FEROS spectra and the average of these same spectra. Gaps in phase coverage are filled in with
grey background color.}
\end{figure}
\clearpage

\begin{figure}[!t]
\includegraphics[width=\columnwidth]{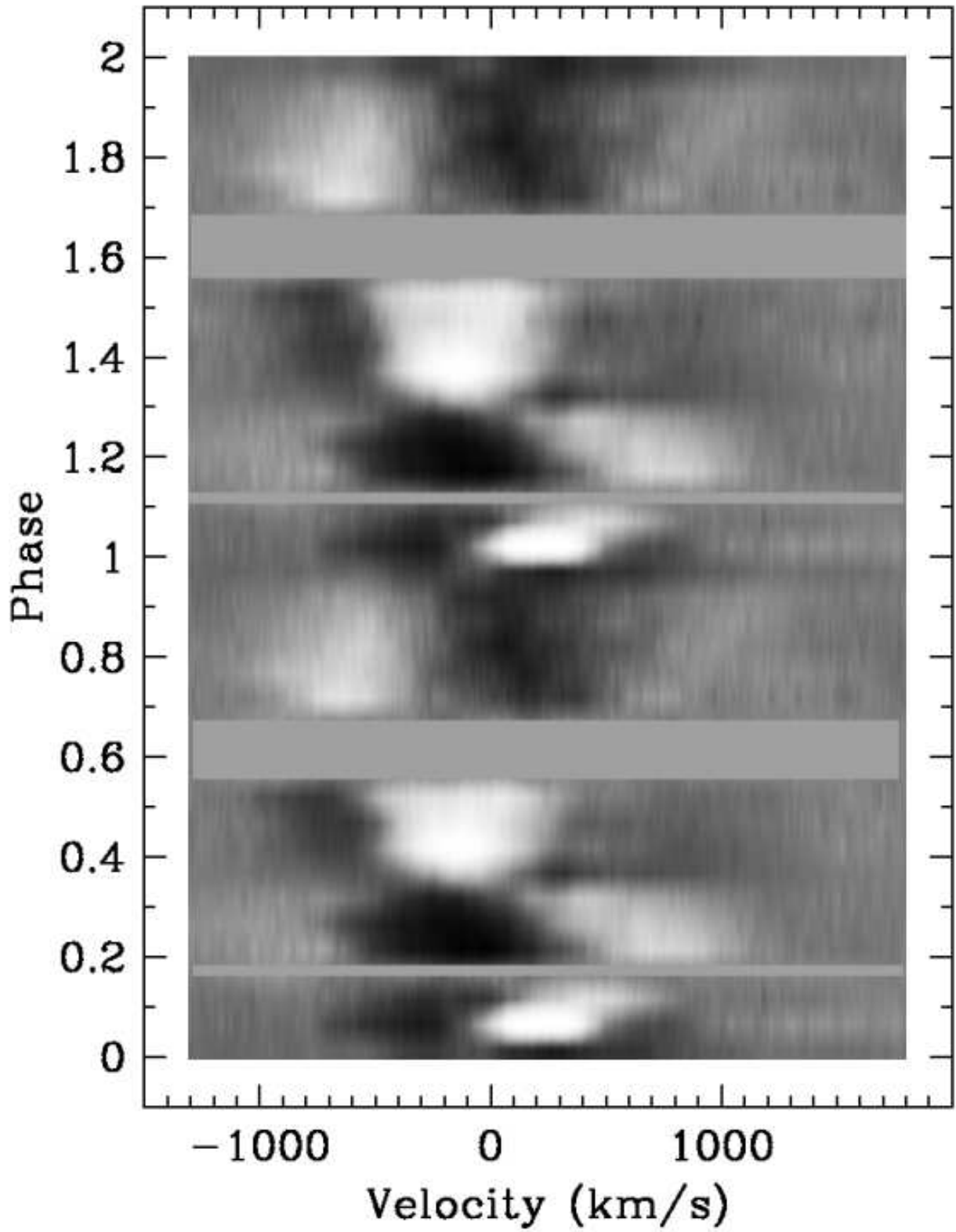}      
\caption{Grey scale representation of 1999 FEROS data for He II 5411 as in
previous figure}
\end{figure}
\clearpage

\begin{figure}[!t]
\includegraphics[width=\columnwidth]{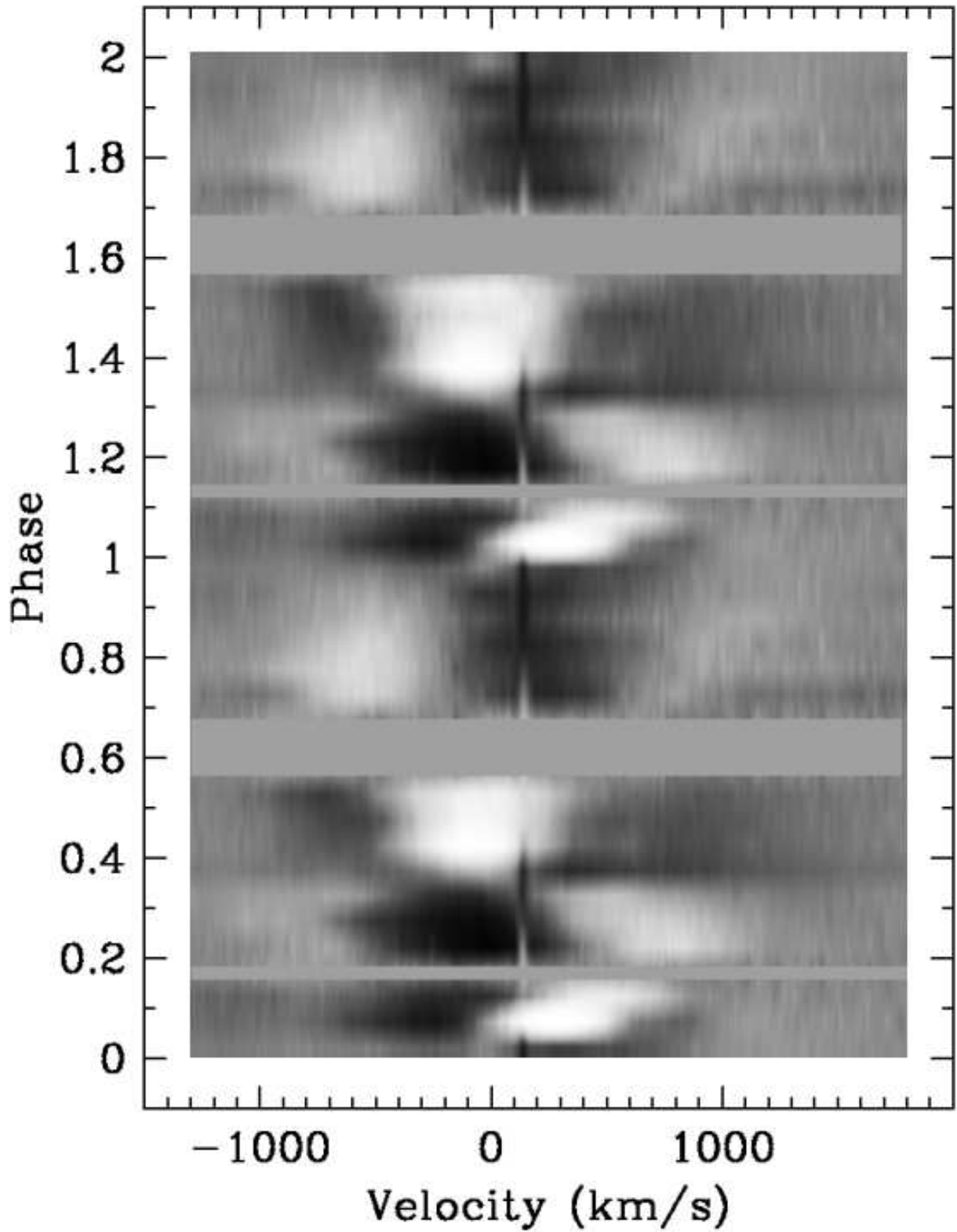}      
\caption{Grey scale representation of 1999 FEROS data for H$\beta$+He II as in previous figure.}
\end{figure}
\clearpage

\begin{figure}[!t]
\includegraphics[width=\columnwidth]{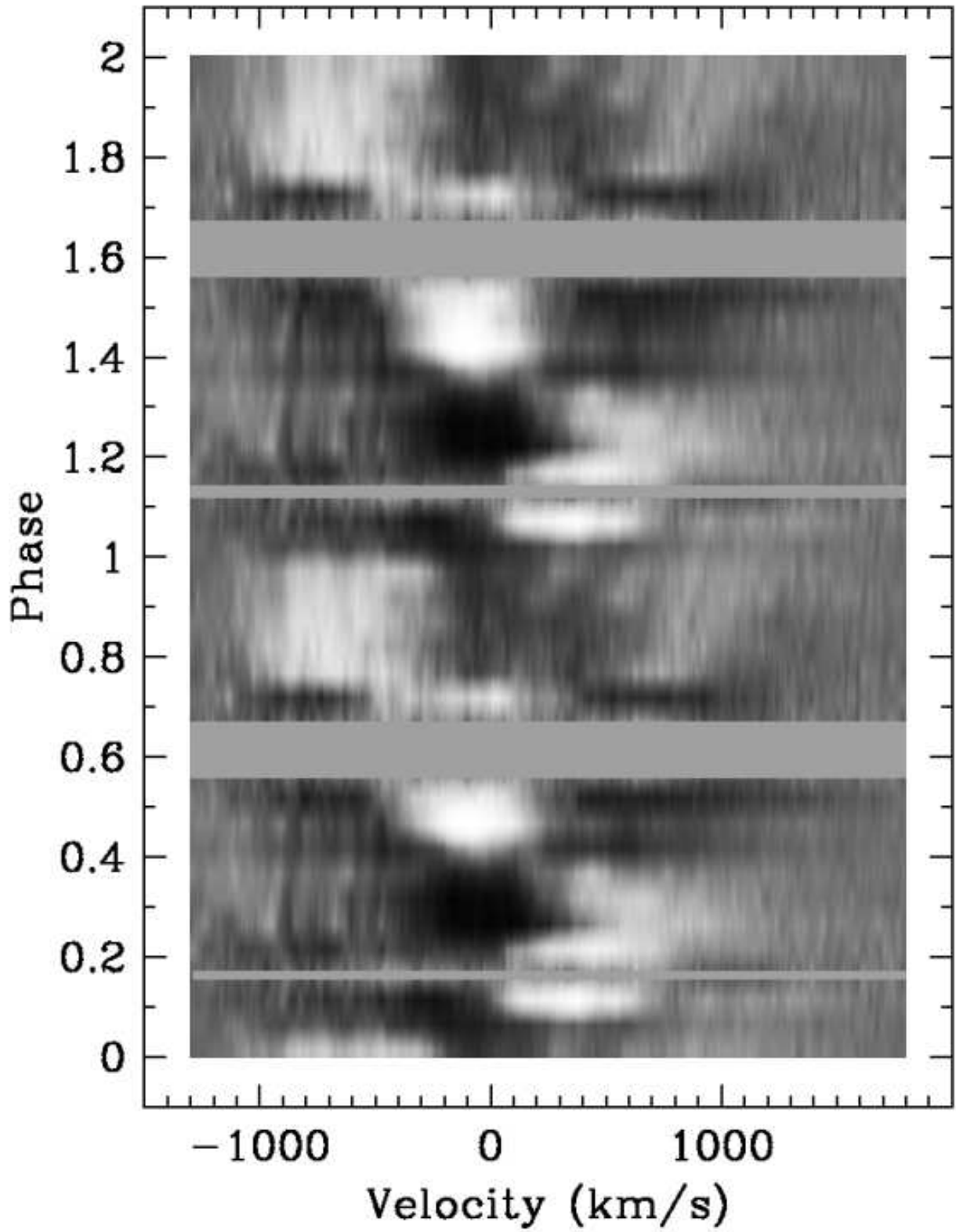}     
\caption{Grey scale representation of 1999 FEROS data for HeI 6678 as in previous figure.}
\end{figure}
\clearpage

\begin{figure}[!t]
\includegraphics[width=\columnwidth]{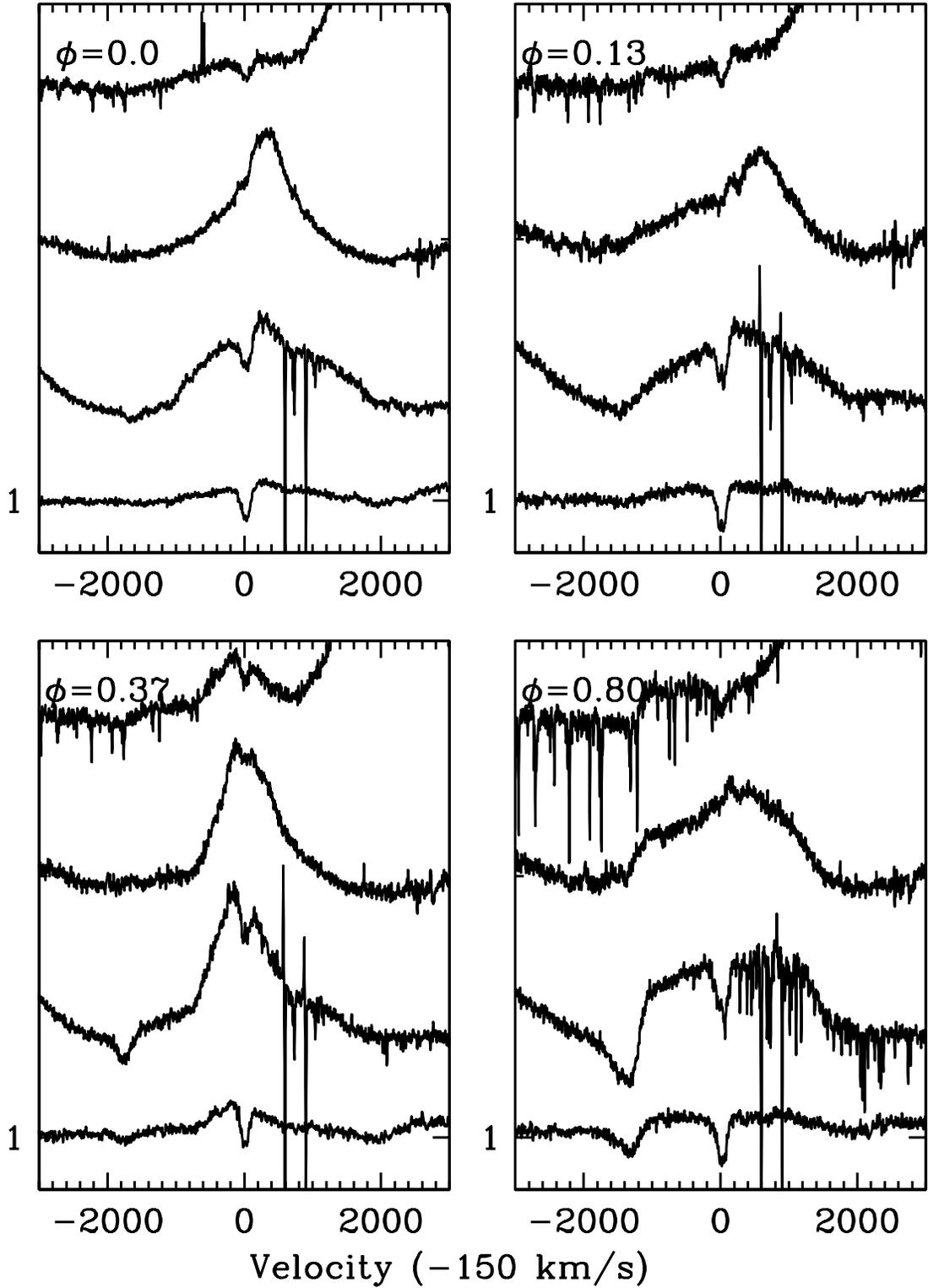}            
\caption{FEROS 2005 line profiles of He I 7065, 6678, 5875 and 4471 \AA\
(from top down). The spectra are averages for each of the orbital phases
$\phi=$0.0 (top left), 0.13 (top right), 0.37 (bottom left) and 0.80 (bottom right). 
At $\phi=$0.0 and 0.37 they are renormalized to account for the  eclipses. The 
velocity scale is corrected for the  +150 km/s systemic velocity of the SMC. Note 
that the photospheric absorption lines are stationary.  The P Cygni absorptions 
are particularly prominent at $\phi=$0.80.
}
\end{figure}
\clearpage

\begin{figure}[!t]
\includegraphics[width=\columnwidth]{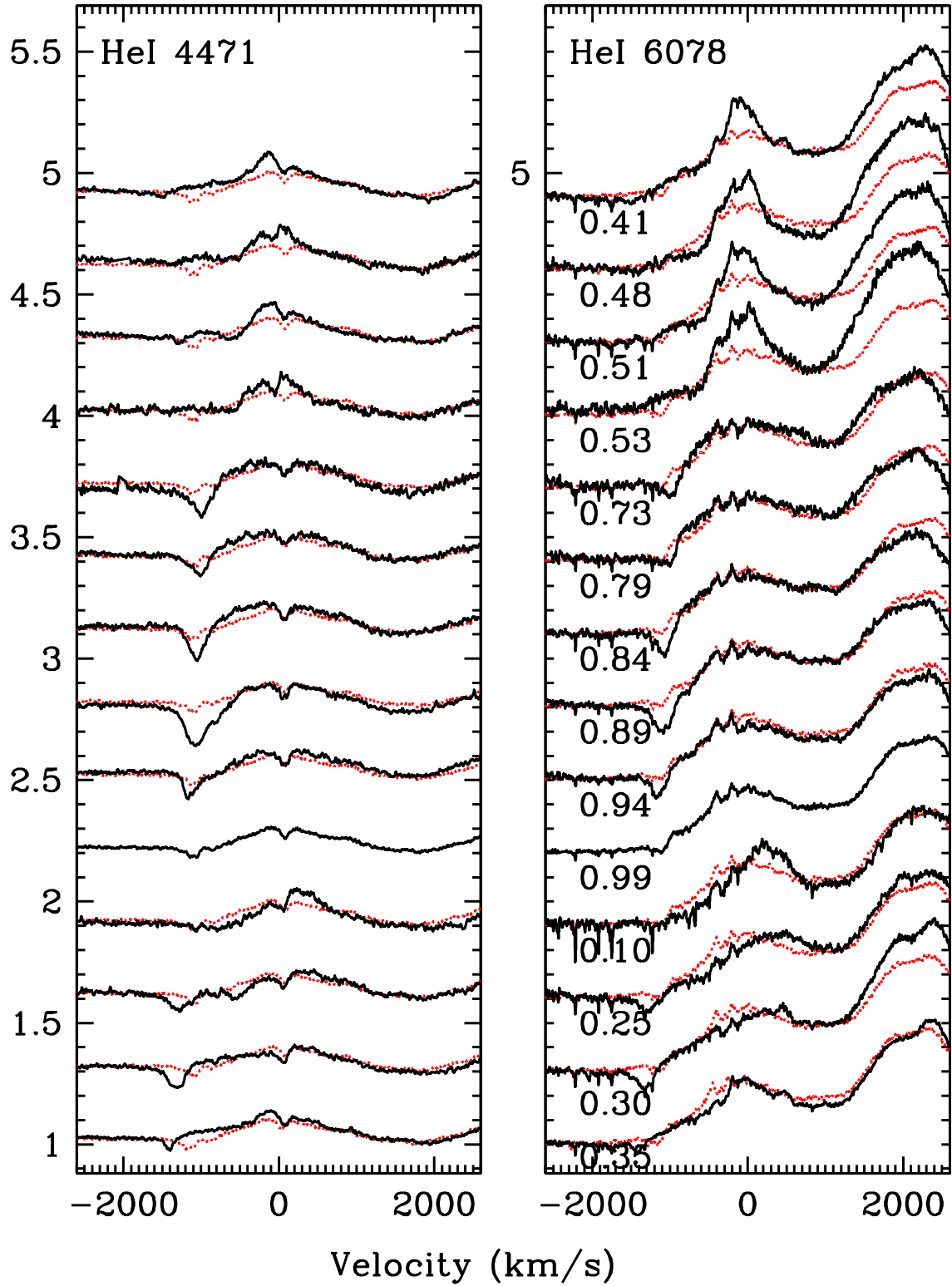}           
\caption{Montage of the 1999 individual FEROS spectra obtained over a single orbital cycle
illustrating the behavior of the He I P Cygni absorption components.  The template
(dotted spectrum) is the profile at $\phi=$0.99, renormalized to account for the eclipse.
The spectrum at $\phi=$0.35 is also renormalized.    The absorption is strong during
the phase interval $\phi=$0.73-0.30, but nearly vanishes at $\phi=$0.99 (eclipse of star B). 
The y-axis is relative intensity, each spectrum having been shifted along this axis by 0.3 units.
The x-axis is velocity with respect to the laboratory wavelength, corrected for SMC systemic
motion (+150 km/s).
}
\end{figure}
\clearpage

\begin{figure}[!t]
\includegraphics[width=\columnwidth]{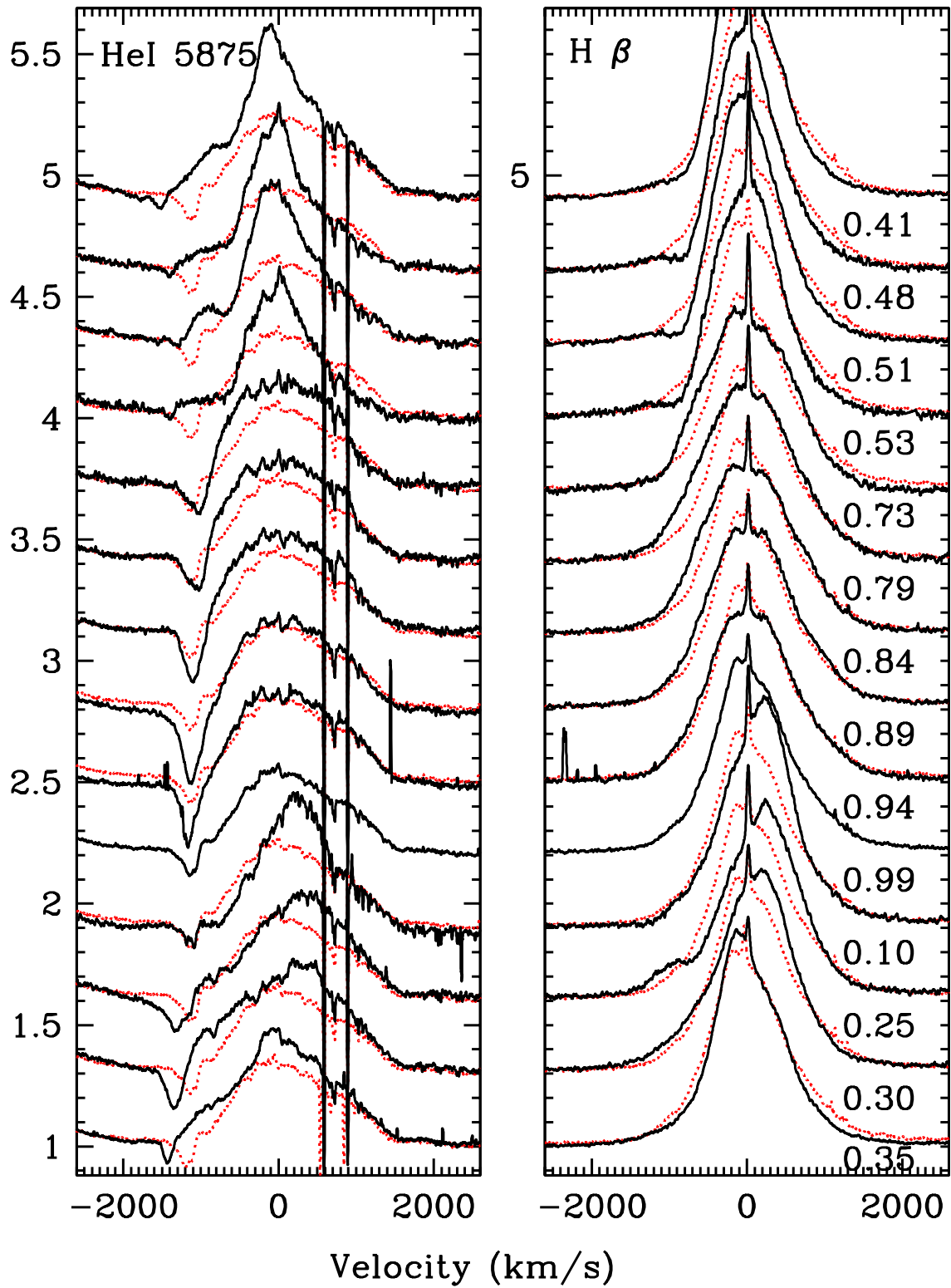}          
\caption{Same as previous figure, but for He I 5875 \AA\ (left).  The same behavior is
observed for the P Cygni absorptions. Note that when they are  sharpest and strongest,
the emission component has a very rounded shape.  When the P Cygni absorption is not
as prominent, the  emission appears to have a sharper (possibly superposed)
additional emission component. The right panel illustrates the same montage of spectra
at H$\beta$.  The y-axis is relative intensity, each spectrum having been shifted
along this axis by 0.3 units. The x-axis is velocity with respect to the laboratory 
wavelength, corrected for SMC systemic motion (+150 km/s).
}
\end{figure}
\clearpage

\begin{figure}[!t]
\includegraphics[width=\columnwidth]{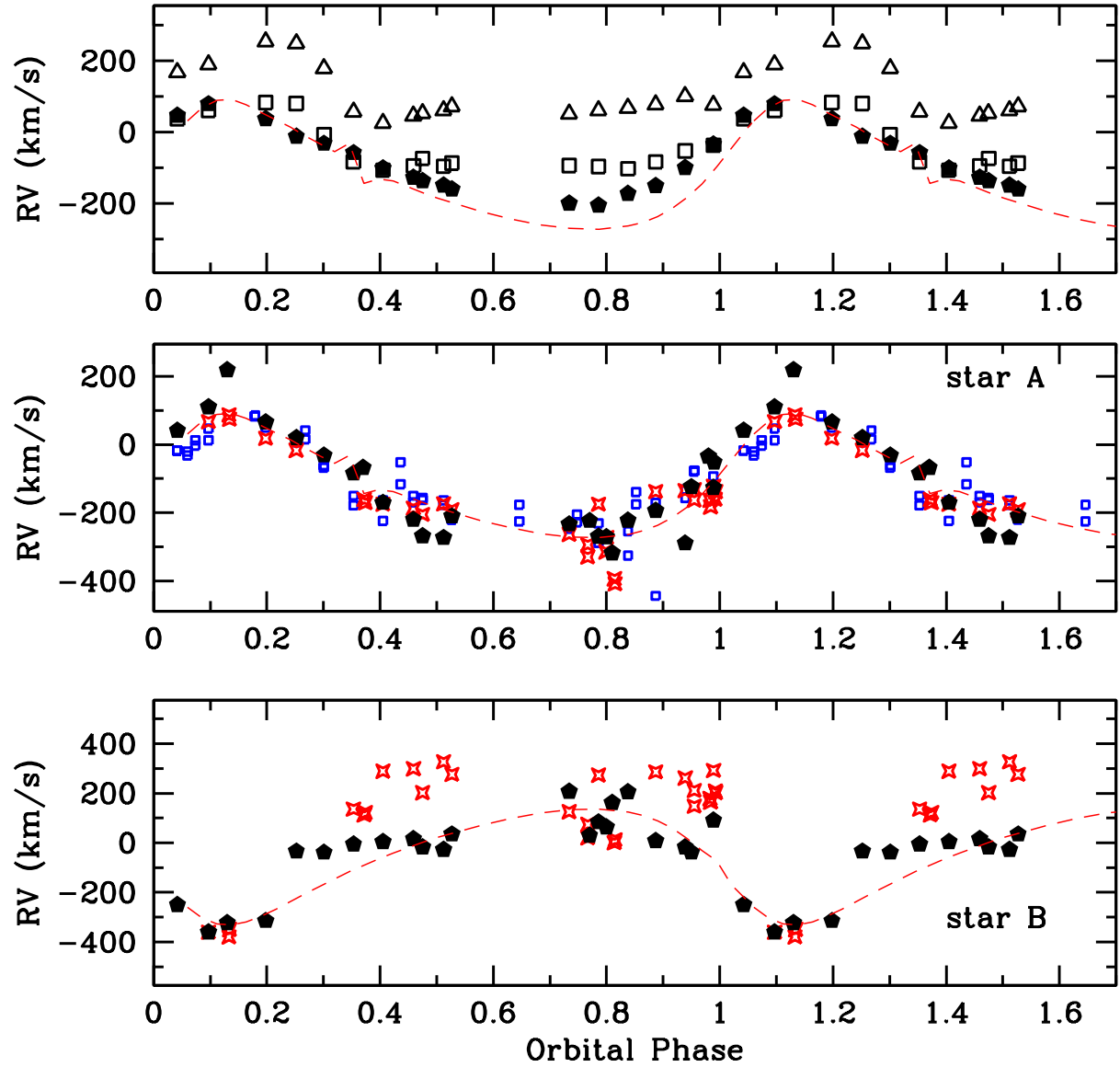}     
\caption{Radial velocity measurements.  Top: A single Gaussian was fit to lines of 
N IV 4057  \AA\ (filled pentagons), H I 4101  \AA\ (squares) and He II 5411  \AA\ (triangles). 
Each RV curve has a different shape due to contributions from different sources. 
Middle:  {\it Star A}'s contribution  isolated using a 2-function fit to  N IV 4057  \AA\ 
(filled pentagons)  and N V 4603  \AA\ (stars), and  a single Gaussian to He II 6074 
and 6118 \AA\ (small squares).  Arbitrary vertical shifts were applied  to correct for 
different zero-point locations and align the 4 RV curves.  Bottom: {\it Star B}'s deduced 
contribution derived from the 2-function fit to N IV 4057  \AA\ (filled pentagons) 
and  N IV 4603  \AA\ (stars). The dashed curves are  theoretical  computations for a 
70$+$54 M$_\odot$ binary system in an e$=$0.3, $\omega=$133$^\circ$ orbit.  All RVs are 
corrected for the SMC systemic velocity (+150 km/s).}
\end{figure}

\begin{figure}[!t]
\includegraphics[width=\columnwidth]{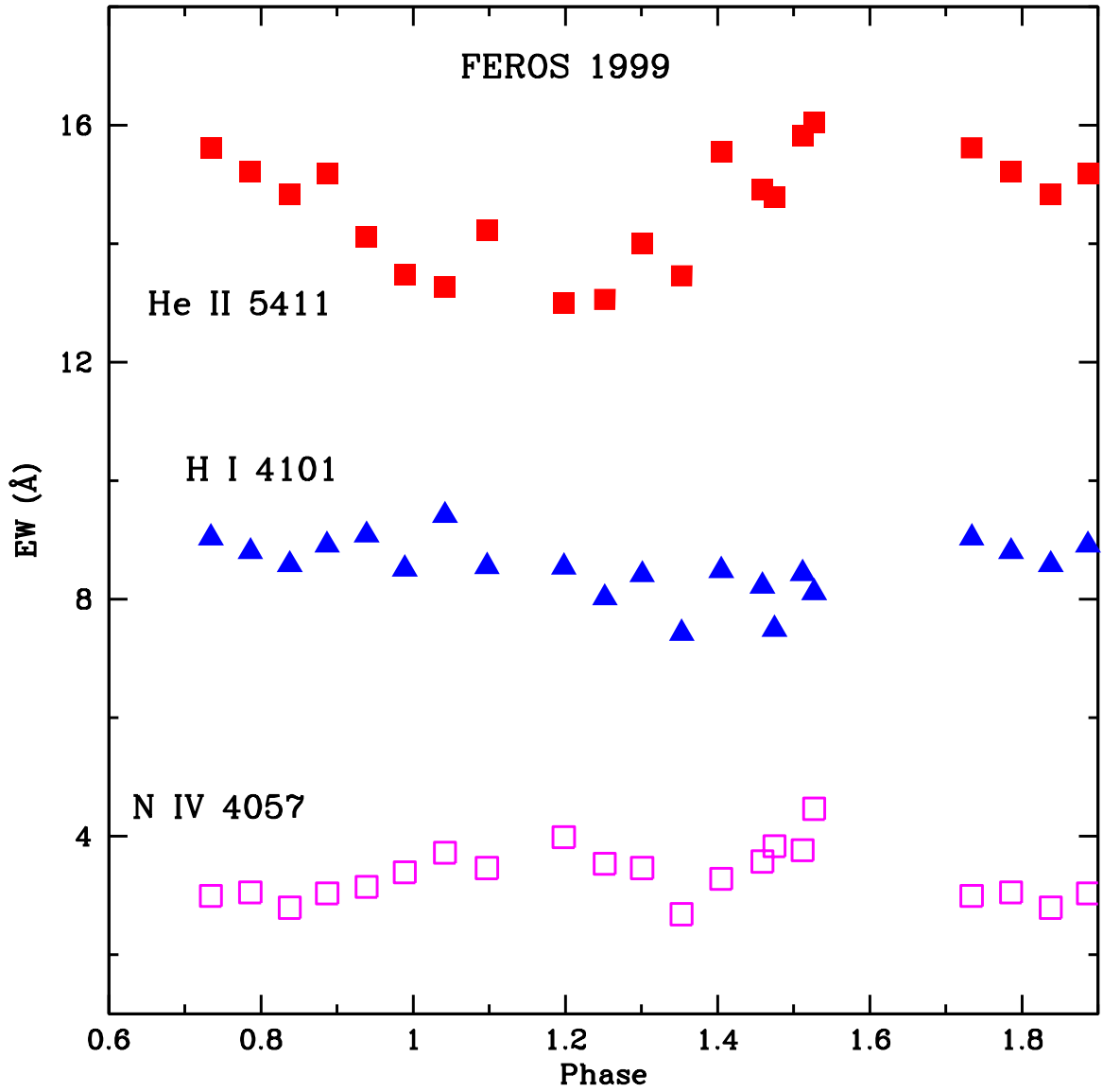}   
\caption{The equivalent width of 1999 FEROS data as a function of orbital phase for
N IV 4057 \AA\, H$\delta+$ He II 4101  \AA\ and He II 5411  \AA. }
\end{figure}

\begin{figure}[!t]
\includegraphics[width=\columnwidth]{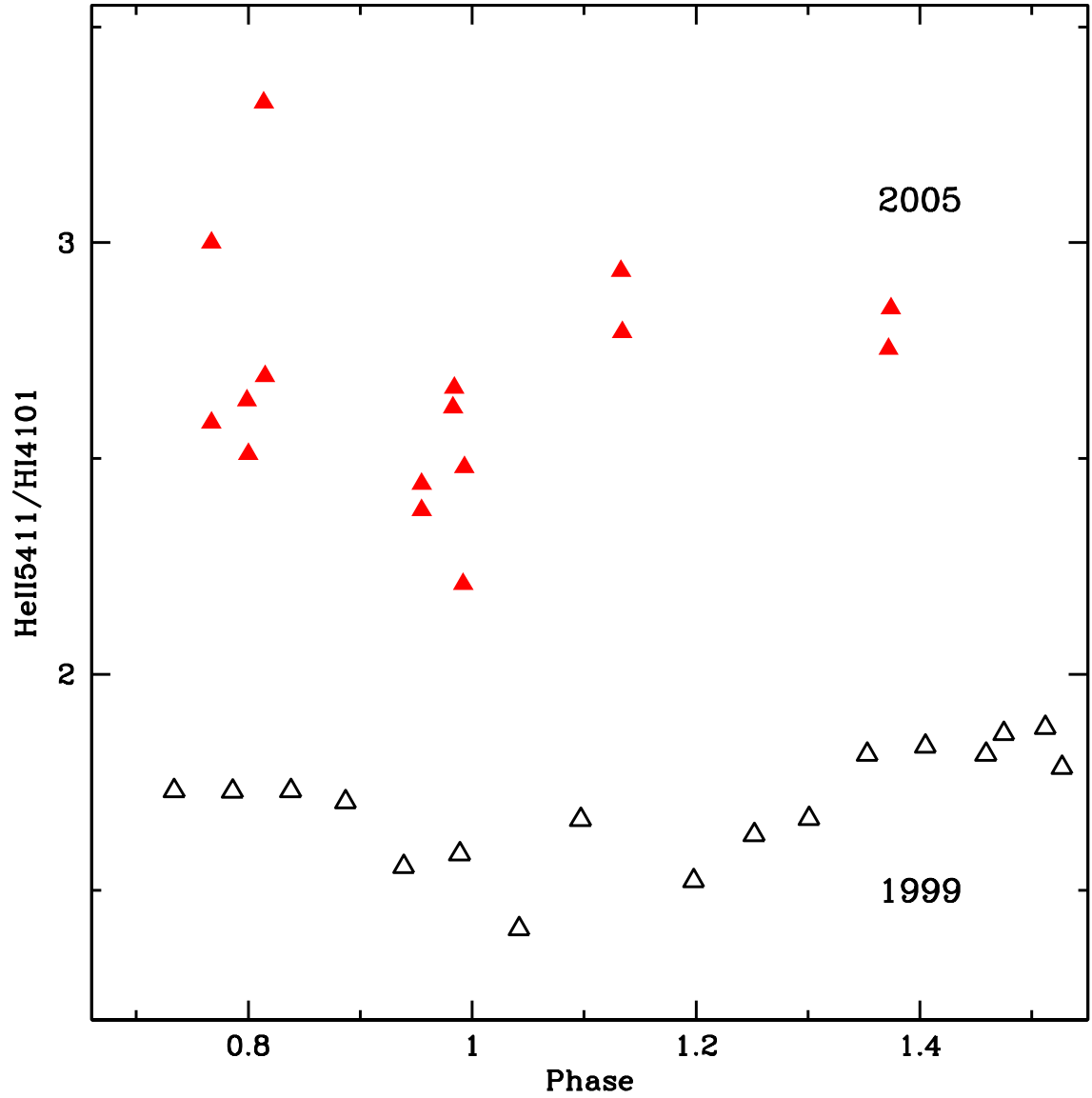}    
\caption{The ratio of equivalent widths of He II 5411 \AA\ and H I (+He II) 4101  \AA\ 
as a function of orbital phase for the  1999 and 2005 FEROS observations.  There is 
only a hint of an orbital-phase dependence in 1999.  Much clearer is the overall 
change in the ratio from 1999 to 2005, indicating that the degree of ionization increased 
in 2005.}
\end{figure}

\begin{figure}[!t]
\includegraphics[width=\columnwidth]{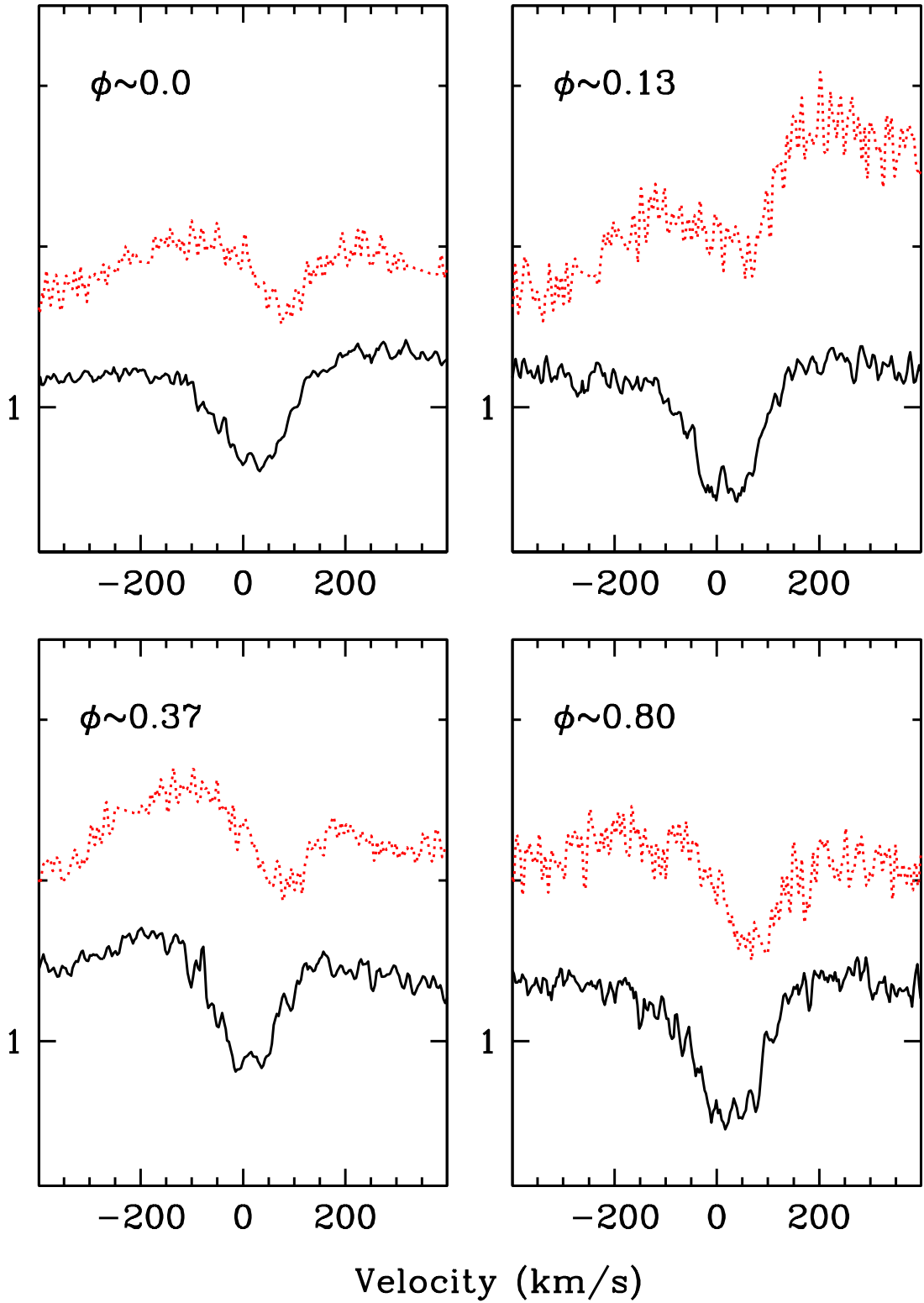}   
\caption{Comparison of the He I 4471 \AA\ photospheric absorption line in
the 1999 (dotted) and in 2005  at four orbital phases.  The velocity
scale is corrected for an assumed SMC systemic velocity of $+$150 km/s, and the
1999 data were shifted upward for clarity in the figure.  Note
that the 2005 profiles are  blue-shifted and stronger than the 1999 profiles.
The 1999  spectra were obtained over the timescale MJD 51381.4-51391.4.
}
\end{figure}
\clearpage

\begin{figure}[!t]
\includegraphics[width=\columnwidth]{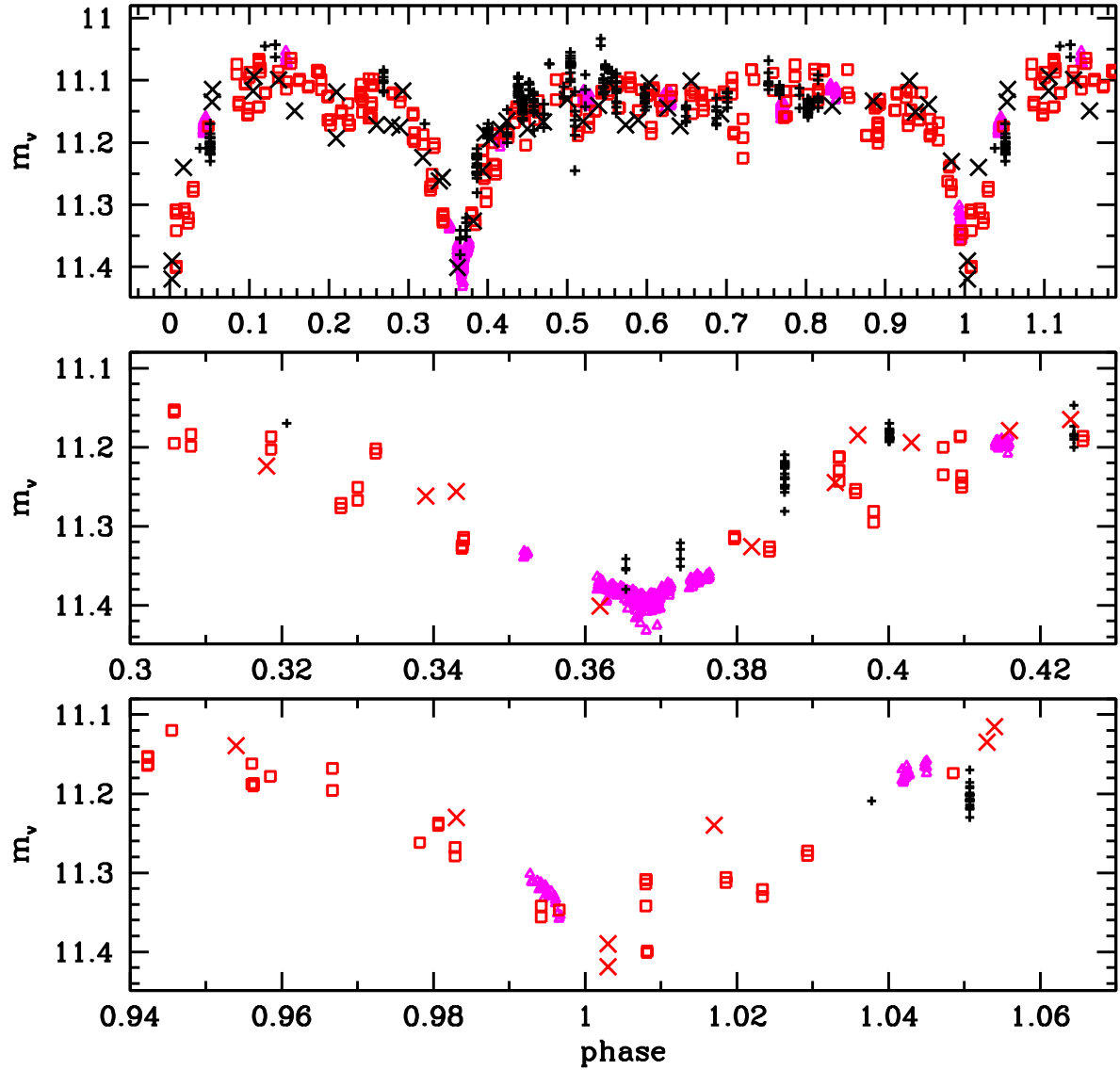}    
\caption{Visual magnitudes from  SMARTS (open squares) and Swope 
(plus signs) compared with  differential photometric observations 
of Breysacher \& Perrier (1980; data of July-October 1979) $+$10.415 mag (crosses); 
and Danish data $+$12.57 mag (filled-in triangles).  Orbital phases for the 
Danish2005, SMARTS and Swope data sets were computed using T$_0=$JD2443158.705 and 
P=19.2654 days from Sterken \& Breysacher (1997). The orbital phases of BP80 data are 
as listed in their Table 1, which BP80 computed with T$_{BP80}=$JD2443158.771 and 
P$_{BP80}=$19.266 days. The difference in phase computed with P and P$_{BP80}$ 
is $<$0.002 for the BP80 data.
}
\end{figure}
\clearpage

\begin{figure}[!t]
\includegraphics[width=\columnwidth]{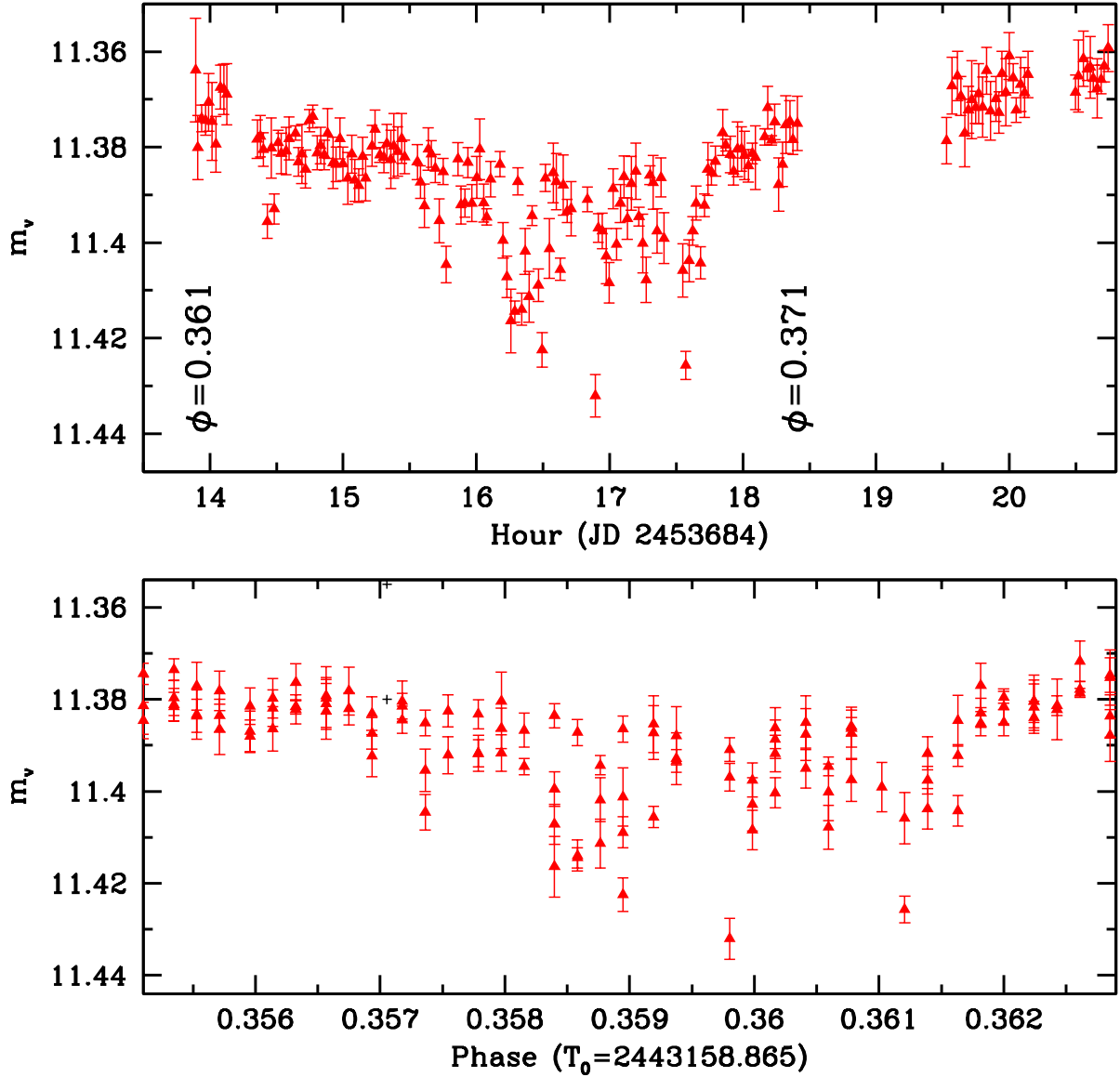}           
\caption{Differential photometric observations from the Danish telescope at 
$\phi=$0.36--0.37, after adding  +12.57 mag to convert to visual magnitudes 
consistent with the SMARTS values. Top: plotted as a function  of time (in hours).  
Bottom: plotted as a function of orbital phase, using P=19.2654 d and 
T$_{modified}=$JD2443158.865. With this T$_{modified}$, the secondary minimum is centered on $\phi=$0.36.  
The error bars correspond to the uncertainty per individual data point, $\sigma_{instr}$.}
\end{figure}
\clearpage

\begin{figure}[!t]
\includegraphics[width=\columnwidth]{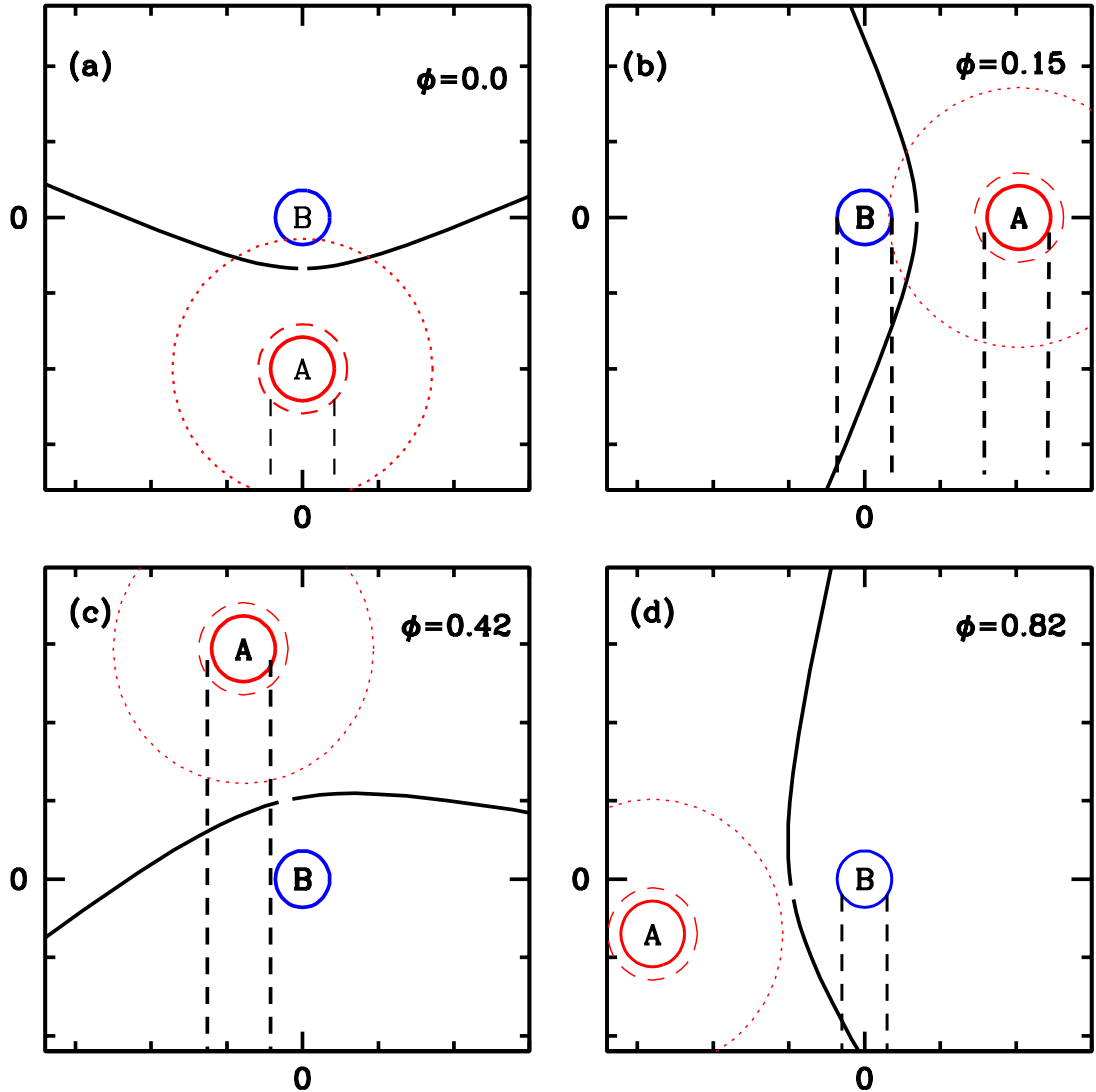}          
\caption{Cartoon of the system geometry for different
orbital phases, illustrating how the lines-of-sight to the two stars
cross different portions of the WWI region and the stellar winds. Dashed
lines contain  the columns of gas projected onto the continuum-emitting regions,
as seen by an observer situated at the bottom of the page. Discontinuous circles
around {\it star A} indicate the accelerating portion of the wind (1.5R$_A$) and
the assumed extent of a line-emitting region ($\sim$4R$_A$). 
}
\end{figure}
\clearpage

\begin{figure}[!t]
\includegraphics[width=\columnwidth]{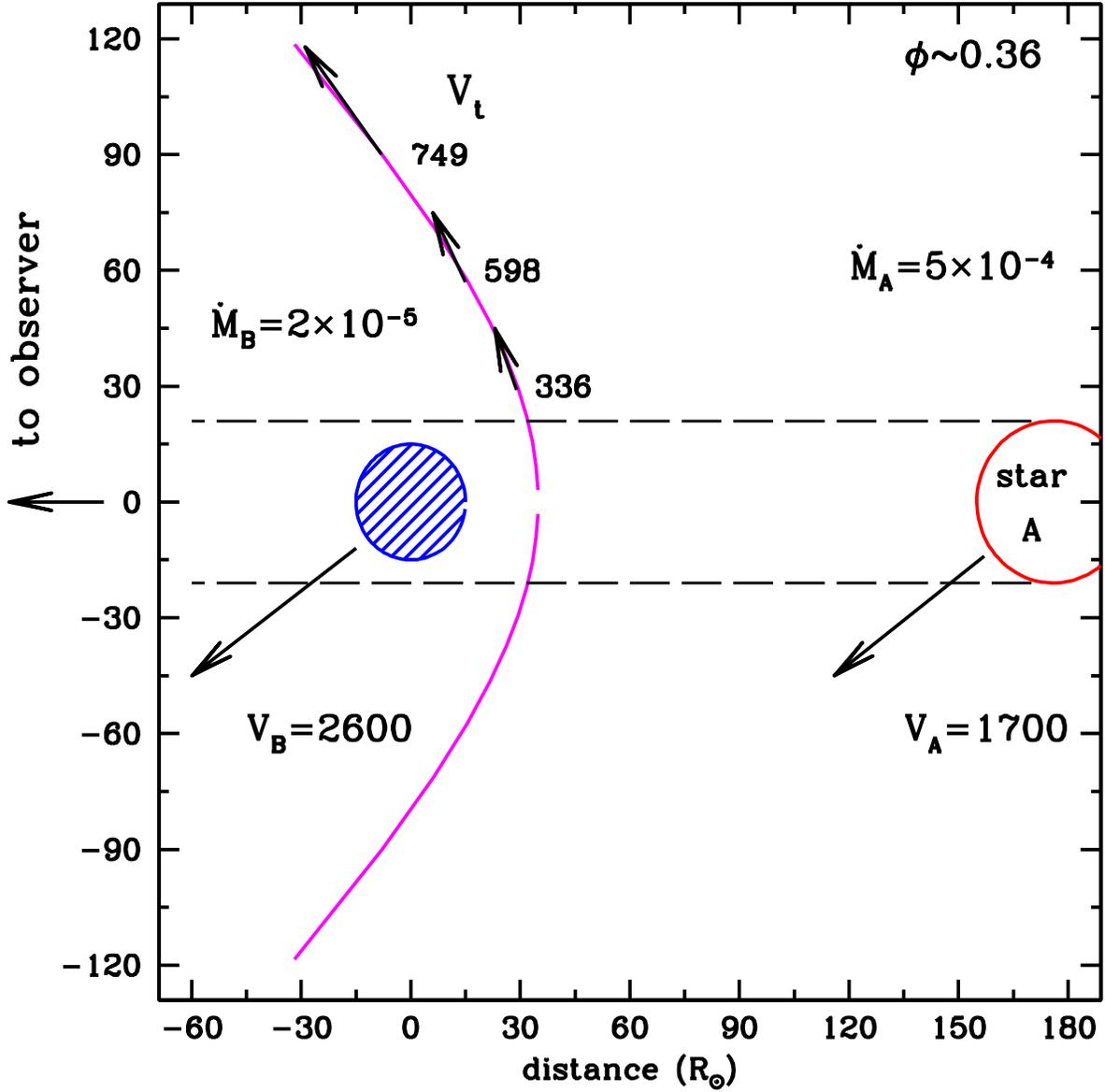}       
\caption{Schematic representation of the WWC shock cone at orbital phase
0.36  listing tangential flow velocities (v$_t$) computed using the relations given by 
Cant\'o et al. (1996).  The assumed mass-loss rates are in M$_\odot$/year and
velocities are in km/s.  The observer is off the page to the left at $\phi=$0.36. The dashed lines
contain the column of material traversed by the line-of-sight from the observer
to {\it star A}.   With the assumed radii (R$_A=$15 R$_\odot$ and R$_B=$21 R$_\odot$),
the light from {\it star A}'s disk must pass through the WWI region {\em and} {\it star B}'s
inner wind before reaching the observer.  Time-dependent mass ejections
from {\it star B} or instabilities along the WWI region could be responsible for the 
rapid variability observed at this orbital phase.
}
\end{figure}

\begin{figure}[!t]
\includegraphics[width=\columnwidth]{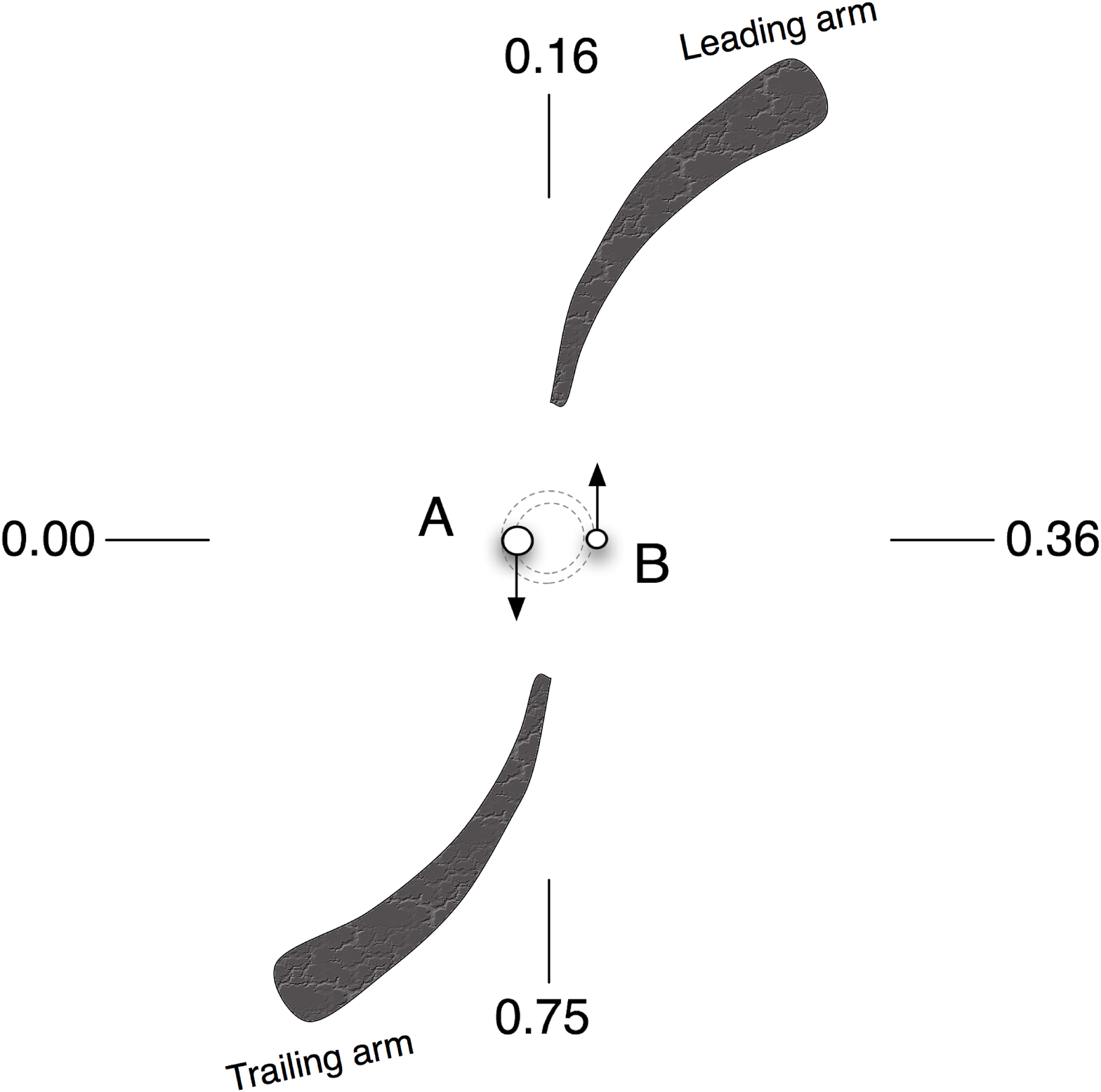}       
\caption{Cartoon illustrating an alternative geometry for the WWI
region that would be consistent with the orbital phase-dependent
behavior of the He I P Cygni absorption components.  
}
\end{figure}
\clearpage

\begin{table*}[!t]\centering
\tablecols{5}
\setlength\tabnotewidth{0.60\textwidth}
\setlength{\tabcolsep}{0.9\tabcolsep}
\small
\caption{Average photometric visual magnitudes: SWOPE and SMARTS}
\begin{tabular}{lrrrr}
\toprule
  $<MJD>$ & $<phase>$ & Num. &   $<m_v>$  & s.d.   \\
\midrule
   52838.379 &    0.464 &   16 &    11.143 &     0.017  \\
   52839.362 &    0.515 &    9 &    11.154 &     0.046  \\
   52840.324 &    0.565 &   12 &    11.123 &     0.019  \\
   52857.303 &    0.446 &   18 &    11.145 &     0.012  \\
   52858.273 &    0.497 &    3 &    11.122 &     0.010  \\
   52859.257 &    0.548 &    5 &    11.086 &     0.018  \\
   52860.244 &    0.599 &    1 &    11.142 &       \nodata  \\
   52972.072 &    0.404 &   15 &    11.185 &     0.006  \\
   52973.061 &    0.455 &    8 &    11.110 &     0.008  \\
   52974.059 &    0.507 &   16 &    11.076 &     0.012  \\
   52975.056 &    0.559 &    4 &    11.090 &     0.005  \\
   52991.107 &    0.392 &   18 &    11.236 &     0.018  \\
   52992.063 &    0.441 &   19 &    11.114 &     0.011  \\
   52992.844 &    0.482 &    2 &    11.073 &     0.001  \\
   52993.905 &    0.537 &    2 &    11.038 &     0.008  \\
   53008.062 &    0.272 &    7 &    11.101 &     0.013  \\
   53009.067 &    0.324 &    1 &    11.170 &       \nodata  \\
   53010.034 &    0.374 &    4 &    11.335 &     0.013  \\
   53011.039 &    0.426 &    6 &    11.180 &     0.018  \\
   53159.403 &    0.127 &    4 &    11.048 &     0.010  \\
   53183.377 &    0.372 &    4 &    11.357 &     0.016  \\
   53210.325 &    0.771 &    7 &    11.117 &     0.005  \\
   53225.305 &    0.548 &    7 &    11.109 &     0.010  \\
   53226.258 &    0.598 &   11 &    11.128 &     0.011  \\
   53227.387 &    0.656 &    6 &    11.154 &     0.012  \\
   53228.295 &    0.703 &    6 &    11.128 &     0.010  \\
   53229.240 &    0.752 &    7 &    11.099 &     0.016  \\
   53230.228 &    0.804 &    8 &    11.135 &     0.011  \\
   53247.271 &    0.688 &    7 &    11.153 &     0.015  \\
   53254.290 &    0.053 &   20 &    11.205 &     0.013  \\
   53288.193 &    0.812 &   26 &    11.137 &     0.015  \\
   53591.385 &    0.550 &    2 &    11.119 &     0.003  \\
   53592.405 &    0.603 &    2 &    11.114 &     0.003  \\
   53593.400 &    0.655 &    1 &    11.120 &       \nodata  \\
   53596.401 &    0.810 &    5 &    11.122 &     0.007  \\
   53597.378 &    0.861 &    1 &    11.128 &       \nodata  \\
   53599.223 &    0.957 &    2 &    11.149 &     0.041  \\
   53600.401 &    0.018 &    2 &    11.325 &     0.006  \\
   53602.355 &    0.119 &    2 &    11.084 &     0.005  \\
   53608.313 &    0.429 &    2 &    11.189 &     0.004  \\
   53615.078 &    0.780 &    1 &    11.158 &       \nodata  \\
   53617.281 &    0.894 &    2 &    11.134 &     0.006  \\
   53618.199 &    0.942 &    3 &    11.138 &     0.016  \\
   53619.232 &    0.995 &    1 &    11.347 &       \nodata  \\
   53620.238 &    0.048 &    1 &    11.174 &       \nodata  \\
   \bottomrule
 \end{tabular}
\end{table*}

\begin{table*}[!t]\centering
\tablecols{5}
\setlength\tabnotewidth{0.60\textwidth}
\setlength{\tabcolsep}{0.9\tabcolsep}
\small
\caption{Average photometric visual magnitudes: SWOPE and SMARTS}
\begin{tabular}{lrrrr}
\toprule
  $<MJD>$ & $<phase>$ & Num. &   $<m_v>$  & s.d.   \\
\midrule
   53621.379 &    0.107 &    2 &    11.091 &     0.006  \\
   53622.366 &    0.158 &    2 &    11.069 &     0.006  \\
   53624.244 &    0.256 &    2 &    11.104 &     0.008  \\
   53625.326 &    0.312 &    2 &    11.191 &     0.011  \\
   53628.267 &    0.464 &    2 &    11.142 &     0.013  \\
   53631.317 &    0.623 &    2 &    11.146 &     0.004  \\
   53632.323 &    0.675 &    2 &    11.145 &     0.006  \\
   53634.303 &    0.778 &    2 &    11.126 &     0.012  \\
   53638.265 &    0.983 &    2 &    11.274 &     0.008  \\
   53640.255 &    0.087 &    2 &    11.137 &     0.004  \\
   53642.245 &    0.190 &    2 &    11.107 &     0.011  \\
   53643.233 &    0.241 &    2 &    11.141 &     0.014  \\
   53646.201 &    0.395 &    2 &    11.288 &     0.010  \\
   53650.388 &    0.613 &    2 &    11.181 &     0.008  \\
   53651.352 &    0.663 &    1 &    11.154 &       \nodata  \\
   53652.397 &    0.717 &    2 &    11.185 &     0.003  \\
   53655.185 &    0.862 &    2 &    11.101 &     0.026  \\
   53657.186 &    0.966 &    2 &    11.175 &     0.018  \\
   53658.132 &    0.015 &    4 &    11.318 &     0.016  \\
   53660.178 &    0.121 &    4 &    11.070 &     0.003  \\
   53663.209 &    0.278 &    2 &    11.121 &     0.001  \\
   53664.219 &    0.331 &    2 &    11.205 &     0.004  \\
   53665.227 &    0.383 &    2 &    11.329 &     0.004  \\
   53666.232 &    0.435 &    2 &    11.144 &     0.003  \\
   53669.062 &    0.582 &    4 &    11.104 &     0.006  \\
   53670.051 &    0.633 &    2 &    11.121 &     0.007  \\
   53672.047 &    0.737 &    1 &    11.099 &       \nodata  \\
   53673.020 &    0.787 &    3 &    11.095 &     0.021  \\
   53675.047 &    0.893 &    2 &    11.139 &     0.002  \\
   53676.180 &    0.951 &    4 &    11.158 &     0.006  \\
   53677.196 &    0.004 &    2 &    11.349 &     0.010  \\
   53679.157 &    0.106 &    2 &    11.150 &     0.007  \\
   53680.269 &    0.164 &    2 &    11.100 &     0.002  \\
   53682.075 &    0.257 &    4 &    11.136 &     0.022  \\
   53683.264 &    0.319 &    2 &    11.195 &     0.011  \\
   53685.105 &    0.415 &    2 &    11.186 &     0.001  \\
   53688.131 &    0.572 &    2 &    11.145 &     0.001  \\
   53689.107 &    0.622 &    2 &    11.133 &     0.000  \\
   53691.197 &    0.731 &    1 &    11.225 &       \nodata  \\
   53692.077 &    0.777 &    2 &    11.159 &     0.001  \\
   53694.082 &    0.881 &    2 &    11.189 &     0.001  \\
   53695.082 &    0.933 &    2 &    11.168 &     0.008  \\
   53698.036 &    0.086 &    2 &    11.082 &     0.011  \\
   53700.083 &    0.192 &    2 &    11.087 &     0.001  \\
   53701.047 &    0.242 &    2 &    11.120 &     0.002  \\
   53703.075 &    0.347 &    2 &    11.326 &     0.002  \\
   \bottomrule
 \end{tabular}
\end{table*}

\begin{table*}[!t]\centering
\tablecols{5}
\setlength\tabnotewidth{0.60\textwidth}
\setlength{\tabcolsep}{0.9\tabcolsep}
\small
\caption{Average photometric visual magnitudes: SWOPE and SMARTS}
\begin{tabular}{lrrrr}
\toprule
  $<MJD>$ & $<phase>$ & Num. &   $<m_v>$  & s.d.   \\
\midrule
   53704.072 &    0.399 &    2 &    11.255 &     0.004  \\
   53706.066 &    0.503 &    2 &    11.109 &     0.017  \\
   53707.092 &    0.556 &    2 &    11.120 &     0.004  \\
   53710.030 &    0.709 &    3 &    11.109 &     0.013  \\
   53711.043 &    0.761 &    2 &    11.102 &     0.021  \\
   53712.096 &    0.816 &    2 &    11.086 &     0.006  \\
   53714.134 &    0.922 &    2 &    11.121 &     0.001  \\
   53715.171 &    0.975 &    2 &    11.182 &     0.020  \\
   53716.027 &    0.020 &    2 &    11.309 &     0.004  \\
   53719.040 &    0.176 &    2 &    11.110 &     0.001  \\
   53720.049 &    0.229 &    2 &    11.171 &     0.001  \\
   53722.022 &    0.331 &    2 &    11.259 &     0.011  \\
   53724.045 &    0.436 &    2 &    11.160 &     0.011  \\
   53725.024 &    0.487 &    2 &    11.115 &     0.023  \\
   53727.029 &    0.591 &    2 &    11.149 &     0.000  \\
   53752.092 &    0.892 &    2 &    11.142 &     0.005  \\
   53756.059 &    0.098 &    2 &    11.105 &     0.001  \\
   53757.027 &    0.148 &    2 &    11.108 &     0.008  \\
   53758.025 &    0.200 &    4 &    11.135 &     0.019  \\
   53759.027 &    0.252 &    3 &    11.138 &     0.035  \\
   53958.306 &    0.596 &    1 &    11.115 &       \nodata  \\
   53960.339 &    0.701 &    2 &    11.120 &     0.004  \\
   53965.295 &    0.958 &    2 &    11.188 &     0.002  \\
   53966.310 &    0.011 &    2 &    11.400 &     0.001  \\
   53968.322 &    0.116 &    2 &    11.143 &     0.001  \\
   53970.319 &    0.219 &    2 &    11.151 &     0.003  \\
   53976.356 &    0.533 &    2 &    11.169 &     0.009  \\
   53983.365 &    0.896 &    9 &    11.183 &     0.011  \\
   53987.332 &    0.102 &    2 &    11.155 &     0.001  \\
   53989.342 &    0.207 &    2 &    11.168 &     0.006  \\
   53991.336 &    0.310 &    3 &    11.168 &     0.023  \\
   53993.339 &    0.414 &    3 &    11.244 &     0.008  \\
   53995.325 &    0.517 &    2 &    11.186 &     0.004  \\
   53998.274 &    0.670 &    2 &    11.126 &     0.008  \\
   53999.263 &    0.722 &    2 &    11.177 &     0.021  \\
   54004.245 &    0.980 &    2 &    11.239 &     0.002  \\
   54007.391 &    0.143 &    2 &    11.091 &     0.008  \\
   54009.379 &    0.247 &    2 &    11.116 &     0.014  \\
   54011.350 &    0.349 &    2 &    11.316 &     0.003  \\
   54013.302 &    0.450 &    2 &    11.177 &     0.039  \\
   54017.288 &    0.657 &    2 &    11.120 &     0.007  \\
   54051.014 &    0.408 &    2 &    11.217 &     0.025  \\
   54052.012 &    0.460 &    2 &    11.155 &     0.011  \\
   54053.041 &    0.513 &    2 &    11.140 &     0.012  \\
   54054.033 &    0.564 &    2 &    11.106 &     0.006  \\
   54056.247 &    0.679 &    1 &    11.124 &       \nodata  \\
   \bottomrule
 \end{tabular}
\end{table*}

\begin{table*}[!t]\centering
\tablecols{5}
\setlength\tabnotewidth{0.60\textwidth}
\setlength{\tabcolsep}{0.9\tabcolsep}
\small
\caption{Average photometric visual magnitudes: SWOPE and SMARTS}
\begin{tabular}{lrrrr}
\toprule
  $<MJD>$ & $<phase>$ & Num. &   $<m_v>$  & s.d.   \\
\midrule
   54057.223 &    0.730 &    2 &    11.083 &     0.000  \\
   54059.177 &    0.832 &    2 &    11.084 &     0.001  \\
   54061.123 &    0.933 &    2 &    11.122 &     0.008  \\
   54062.195 &    0.988 &    1 &    11.262 &       \nodata  \\
   54066.001 &    0.186 &    4 &    11.085 &     0.001  \\
   54068.153 &    0.297 &    2 &    11.130 &     0.007  \\
   54070.125 &    0.400 &    4 &    11.224 &     0.015  \\
   54074.173 &    0.610 &    2 &    11.124 &     0.001  \\
   54080.161 &    0.921 &    2 &    11.148 &     0.001  \\
   54082.233 &    0.028 &    2 &    11.275 &     0.004  \\
   54084.189 &    0.130 &    2 &    11.120 &     0.001  \\
   54086.101 &    0.229 &    2 &    11.151 &     0.019  \\
   54088.193 &    0.338 &    2 &    11.274 &     0.004  \\
   54089.195 &    0.390 &    2 &    11.315 &     0.002  \\
   54090.200 &    0.442 &    2 &    11.156 &     0.001  \\
   54092.199 &    0.546 &    1 &    11.150 &       \nodata  \\
   \bottomrule
 \end{tabular}
\end{table*}


\end{document}